\documentclass[twocolumn]{aastex63}
\usepackage{multirow}
\usepackage[figuresright]{rotating}
\usepackage{subfigure}
\usepackage{epic,eepic}
\usepackage{graphicx}
\usepackage{longtable}
\usepackage{float}
\usepackage{lineno}
\usepackage{pifont}
\usepackage{gensymb}
\usepackage{multirow}
\usepackage{amsmath}
\usepackage{color}

\shorttitle{}
\shortauthors{Zhao et al.}

\begin{document}

\title{Wide binaries with white dwarf or neutron star companions discovered from Gaia DR3 and LAMOST}

\correspondingauthor{Song Wang}
\email{songw@bao.ac.cn}

\author{Xinlin Zhao}
\affiliation{Key Laboratory of Optical Astronomy, National Astronomical Observatories, Chinese Academy of Sciences, Beijing 100101, China}
\affiliation{College of Astronomy and Space Sciences, University of Chinese Academy of Sciences, Beijing 100049, China}

\author{Huijun Mu}
\affiliation{International Laboratory for Quantum Functional Materials of Henan, and School of Physics and Microelectronics, Zhengzhou University, Zhengzhou, Henan 450001, China}

\author{Song Wang}
\affiliation{Key Laboratory of Optical Astronomy, National Astronomical Observatories, Chinese Academy of Sciences, Beijing 100101, China}
\affiliation{Institute for Frontiers in Astronomy and Astrophysics, Beijing Normal University, Beijing 102206, China}

\author{Xue Li}
\affiliation{Key Laboratory of Optical Astronomy, National Astronomical Observatories, Chinese Academy of Sciences, Beijing 100101, China}
\affiliation{College of Astronomy and Space Sciences, University of Chinese Academy of Sciences, Beijing 100049, China}

\author{Junhui Liu}
\affiliation{Department of Astronomy, Xiamen University, Xiamen, Fujian 361005, People's Republic of China}

\author{Bowen Huang}
\affiliation{Department of Astronomy, Beijing Normal University No. 19, XinJieKouWai St, Beijing 100875, China}

\author{Weimin Gu}
\affiliation{Department of Astronomy, Xiamen University, Xiamen, Fujian 361005, People's Republic of China}

\author{Junfeng Wang}
\affiliation{Department of Astronomy, Xiamen University, Xiamen, Fujian 361005, People's Republic of China}

\author{Tuan Yi}
\affiliation{Department of Astronomy, Xiamen University, Xiamen, Fujian 361005, People's Republic of China}

\author{Zhixiang Zhang}
\affiliation{Department of Astronomy, Xiamen University, Xiamen, Fujian 361005, People's Republic of China}

\author{Haibo Yuan}
\affiliation{Department of Astronomy, Beijing Normal University No. 19, XinJieKouWai St, Beijing 100875, China}
\affiliation{Institute for Frontiers in Astronomy and Astrophysics, Beijing Normal University, Beijing 102206, China}

\author{Zhongrui Bai}
\affiliation{Key Laboratory of Optical Astronomy, National Astronomical Observatories, Chinese Academy of Sciences, Beijing 100101, China}

\author{Hailong Yuan}
\affiliation{Key Laboratory of Optical Astronomy, National Astronomical Observatories, Chinese Academy of Sciences, Beijing 100101, China}

\author{Haotong Zhang}
\affiliation{Key Laboratory of Optical Astronomy, National Astronomical Observatories, Chinese Academy of Sciences, Beijing 100101, China}

\author{Jifeng Liu}
\affiliation{Key Laboratory of Optical Astronomy, National Astronomical Observatories, Chinese Academy of Sciences, Beijing 100101, China}
\affiliation{College of Astronomy and Space Sciences, University of Chinese Academy of Sciences, Beijing 100049, China}
\affiliation{Institute for Frontiers in Astronomy and Astrophysics, Beijing Normal University, Beijing 102206, China}
\affiliation{WHU-NAOC Joint Center for Astronomy, Wuhan University, Wuhan, Hubei 430072, China}

\begin{abstract}

{\it Gaia} DR3 mission has identified and provided about 440,000 binary systems with orbital solutions, offering a valuable resource for searching binaries including a compact component. By combining the {\it Gaia} DR3 data with radial velocities (RVs) from the LAMOST spectroscopic survey, we identify three wide binaries possibly containing a compact object. For two of these sources with a main-sequence companion, no obvious excess is observed in the blue/red band of the {\it Gaia} DR3 XP spectra, and the LAMOST medium-resolution spectra exhibit clear single-lined features. The absence of an additional component from spectral disentangling analysis further suggests the presence of compact objects within these systems. On the other hand, the visible star of the third source is a stripped giant star. In contrast to most binaries including stripped stars, no emission line is detected in the optical spectra. The unseen star could potentially be a massive white dwarf or neutron star, but the possibility of an F-type dwarf star scenario cannot be ruled out. An examination of about ten binaries containing white dwarfs or neutron stars using both kinematic and chemical methods suggest most of these systems are located in the thin disk of the Milky Way.

\end{abstract}


\section{INTRODUCTION}
\label{intro.sec}

The evolution path a star takes in its life depends on its initial mass, metallicity, and rotational velocity, etc.
As the final product of stellar evolution, compact objects are classified into three categories: white dwarfs, neutron stars, and black holes. 
In the era of multi-messenger astronomy, several techniques have been developed to search and identify compact objects, including X-ray observations \citep{2006ARA&A..44...49R}, gravitational wave detections \citep{2016PhRvL.116f1102A,2017PhRvL.119p1101A}, gravitational microlensing \citep{2022ApJ...933L..23L,2022ApJ...933...83S}, radial velocity (RV) monitoring\citep{2014Natur.505..378C,2019Natur.575..618L,2019Sci...366..637T}, and astrometry \citep{2023MNRAS.518.1057E,2023MNRAS.521.4323E}.
Each method possesses its own strengths and limitations. 
For instance, the X-ray method can be used to detect close binaries with strong accretion, whereas the gravitational wave method is limited to binaries with two compact objects. 
The radial velocity method is well-suited for the search of quiescent compact objects in binaries, while the astrometric method is more suitable for identifying compact objects in long-period binaries.
Recently, a group of black holes and neutron stars have been discovered through RV monitoring with large spectroscopic surveys \citep[e.g.,][]{2019Natur.575..618L,2019Sci...366..637T,2020A&A...637L...3R,2021MNRAS.504.2577J,2022NatAs...6.1203Y,2022ApJ...940..165Y}, although some systems are still in debate.

{\it Gaia} spacecraft is a mission designed to explore the Milky Way by observing over a billion stars, utilizing its astrometry, photometry, and spectroscopy capabilities.
The spacecraft is equipped with three main instruments \citep{2022arXiv220800211G}: an astrometric instrument, prism photometers, and a Radial Velocity Spectrometer (RVS). 
The RVS has a resolution of approximately $R \sim$ 11,500, enabling the measurement of radial velocities of bright stars ($G < 12$ mag).
The {\it Gaia} Data Release 3 (DR3) released four non-single star (NSS) tables for 813,687 objects classified as binaries or multiple stars, including {\it $nss\_two\_body\_orbit$}, {\it $nss\_acceleration\_astro$}, {\it $nss\_non\_linear\_spectro$}, and {\it $nss\_vim\_fl$} \citep{2022gdr3.reptE....V}. 
Of particular interest is the {\it $nss\_two\_body\_orbit$}, which provides orbital solutions for astrometric, spectroscopic, and eclipsing binaries, flagged as ``Orbital", ``SB1", ``SB2", ``AstroSpectroSB1", ``EclipsingBinary", etc.
This comprehensive catalogue serves as a rich reservoir for studying stellar multiplicity and analyzing various aspects of binary systems.

Large Sky Area Multi-Object Fiber Spectroscopic Telescope (hereafter LAMOST), also called GuoShouJing Telescope, is a specialized reflecting Schmidt telescope with an effective aperture of 3.6-4.9 m and a field of view of $5^{\circ}$ \citep{1996ApOpt..35.5155W}. 
The focal plane is equipped with 4000 precisely positioned fibers that are connected to 16 spectrographs \citep{2012RAA....12.1197C,2012RAA....12..723Z}. 
From 2011 to 2018, LAMOST conducted its first stage focused on a low-resolution ($R \sim$ 1,800) spectral survey.
The low-resolution spectrum (LRS) covers a wavelength range of 3650--9000 \AA.
Since October 2018, LAMOST started its second 5-year survey program, containing both low- and medium-resolution ($R \sim$ 7500) spectral surveys.
The medium-resolution spectrum (MRS) spans wavelength ranges of 4950--5350 \AA\ and 6300--6800 \AA\ \citep{2020arXiv200507210L}.
The LAMOST Data Release 9 (DR9) has released more than 19.47 million spectra for stars, galaxies, and quasars, etc.
Notably, the time-domain spectral survey in the second stage provides a great opportunity to achieve a breakthrough in various scientific topics, such as binary systems, stellar activity, stellar pulsation, etc \citep{2021RAA....21..292W}.

By combining the optical spectra from LAMOST with orbital solutions from {\it Gaia} DR3, we searched for binaries containing a compact object. 
As a result, we identified three potential candidates, where the visible stars are either giants or main-sequence stars, and the unseen companions are white dwarfs or neutron stars.
In Section \ref{obs.sec}, we provided an overview of the sample selection and data reduction.
Detailed information about the three systems, including stellar parameters of the visible stars, Kepler orbital solutions, and masses of the unseen companions, are presented in Sections \ref{g4031.sec} to \ref{g8441.sec}.
Finally, a short summary is provided in Section \ref{sum.sec}.

\section{Sample selection and data reduction}
\label{obs.sec}

\subsection{Source selection}

The {\it $nss\_two\_body\_orbit$} table includes 443,205 binaries or multiple star systems.
For spectroscopic binaries (flagged as ``SB1", ``SB2", or ``AstroSpectroSB1"), it provides orbital parameters such as period $P$, eccentricity $e$, and semi-amplitude $K$ of primary, etc.
For ``AstroSpectroSB1" binaries, the $K$ value can be calculated from the astrometric solutions ($c\_thiele\_innes$ and $h\_thiele\_innes$).
First, we selected single-lined systems (``SB1" and ``AstroSpectroSB1") and cross-matched them with the LAMOST DR9 low-resolution and med-resolution general catalog\footnote{http://www.lamost.org/dr9/v1.0/catalogue}.
Second, we chose sources with more than two LAMOST spectral observations with signal-to-noise ratio ($SNR$) greater than 5 and clear RV variation. 
Third, we calculated the binary mass function of those systems using orbital parameters from the {\it $nss\_two\_body\_orbit$}, and the gravitational masses of visible stars with multi-band magnitudes and atmospheric parameters from LAMOST parameter catalog. 
Finally, we estimated the minimum masses of the unseen stars using the mass functions and the evolutionary masses of the visible stars.

Following these steps, three binaries with possible compact objects were selected, namely {\it Gaia} ID
4031997035561149824 (R.A. = 176.46978$^o$; Decl. = 35.29075$^o$; hereafter G4031), 3431326755205579264 (R.A. = 90.61903$^o$; Decl. = 28.13287$^o$; hereafter G3431), and 844176650958726144 (R.A. = 169.12464$^o$; Decl. = 55.72840$^o$; hereafter G8441).
The positions on the Hertzsprung–Russell diagram (Figure \ref{HR_diagram.fig}) indicate that the visible stars of G4031 and G3431 are main-sequence stars while the visible star of G8441 is a giant.

\subsection{Data reduction}

We obtained all available LRS and MRS observations from the LAMOST archive for our candidate sources.
These spectra have undergone various reduction steps with the LAMOST 2D pipeline, including bias and dark subtraction, flat field correction, spectrum extraction, sky background subtraction, and wavelength calibration, etc.\citep[See][for detalis]{2015RAA....15.1095L}. 
Figure \ref{HR_diagram.fig} displays the LRS observations for the three targets.
The cross-correlation function (CCF) was used to calculate the RV values using the red band (6300--7000 \AA) of each MRS spectrum with $SNR>5$.
Furthermore, for each spectrum, we derived a calibration factor ($\Delta rv = rv\_r1 - rv\_r0$) with the $rv\_r1$ and $rv\_r0$ from the LAMOST MRS catalog.
Here $rv\_r1$ represents the calibrated velocity of $rv\_r0$ by using RV standard stars \citep{2018AJ....156...90H}.
The final RV values were obtained by summing the measured value from the CCF method and the calibration factors (Table \ref{lamost3.tab}).

In addition, we applied for spectral observations using the Beijing Faint Object Spectrograph and Camera (BFOSC) mounted on the 2.16 m telescope at the Xinglong Observatory and the Double Spectrograph (DBSP) mounted on the Palomar's 200 in. telescope (P200).
For G8441, we carried out eight observations using the 2.16 m telescope from Feb. 20th, 2019 to Apr. 21th, 2019, with the E9/G10 grism and a 1.6$^{\prime\prime}$ slit configuration, and four observations using the P200 telescope on Mar. 14th, 2019.
For G3431, we applied for two observations using the 2.16 m telescope on Oct. 4th, 2022.
The observed spectra were reduced using the IRAF v2.16 software \citep{1986SPIE..627..733T,1993ASPC...52..173T} following standard steps, and the reduced spectra were then corrected to vacuum wavelength.
The CCF method was used to calculate RV, and the barycentric corrections to the observation time and RV were made using the light\_travel\_time and radial\_velocity\_correction functions provided by the Python package Astropy \citep{2013A&A...558A..33A,2018AJ....156..123A,2022ApJ...935..167A}.

\section{G4031}
\label{g4031.sec}

\subsection{Stellar parameters of the visible star}

Each of our targets has been observed multiple times by LAMOST.
Their atmospheric parameters can be estimated following \citep{2020ApJS..251...15Z}:
\begin{equation} \label{eq1}
\overline{P} = \frac{\sum_k w_k \cdot P_{k}}{\sum_k w_k}
\end{equation}
and
\begin{equation} \label{eq2}
\sigma_w(\overline{P}) = \sqrt{\frac{N}{N-1}\frac{\sum_k w_k \cdot (P_{k} - \overline{P})^2}{\sum_k w_k}},
\end{equation}
where the index $k$ is the $k_{th}$ epoch of the measurements of parameter $P$ (i.e., $T_{\rm eff}$, log$g$, and [Fe/H]) for each star, and the weight $w_k$ is square of {\it SNR} of each spectrum corresponding to the $k_{th}$ epoch.
Table \ref{atmo_pars.tab} lists stellar parameters of G4031 measured by different methods, such as the LAMOST Stellar Parameter Pipeline (LASP) \citep{2015RAA....15.1095L}, DD-Payne \citep{2019ApJS..245...34X,2019ApJ...879...69T}, and CNN \citep{2020ApJ...891...23W}. 
In the following analysis, we preferred to use the LASP LRS results followed by the LASP MRS results.
The stellar parameters of G4031 are $T_{\rm eff} = 5953{\pm 17} K$, log$g$ $= 4.05{\pm 0.02}$ dex, and [Fe/H] $= -0.37{\pm 0.01}$ (Table \ref{stellar_parameters.tab}).

The distance from {\it Gaia} DR3 is approximately 720 pc. 
The Pan-STARRS DR1 3D dust map\footnote{http://argonaut.skymaps.info} does not provide an extinction estimation, instead we used the SFD value with $E(B-V) =0.02$.

\begin{figure*}
    \center
    \includegraphics[width=1\textwidth]{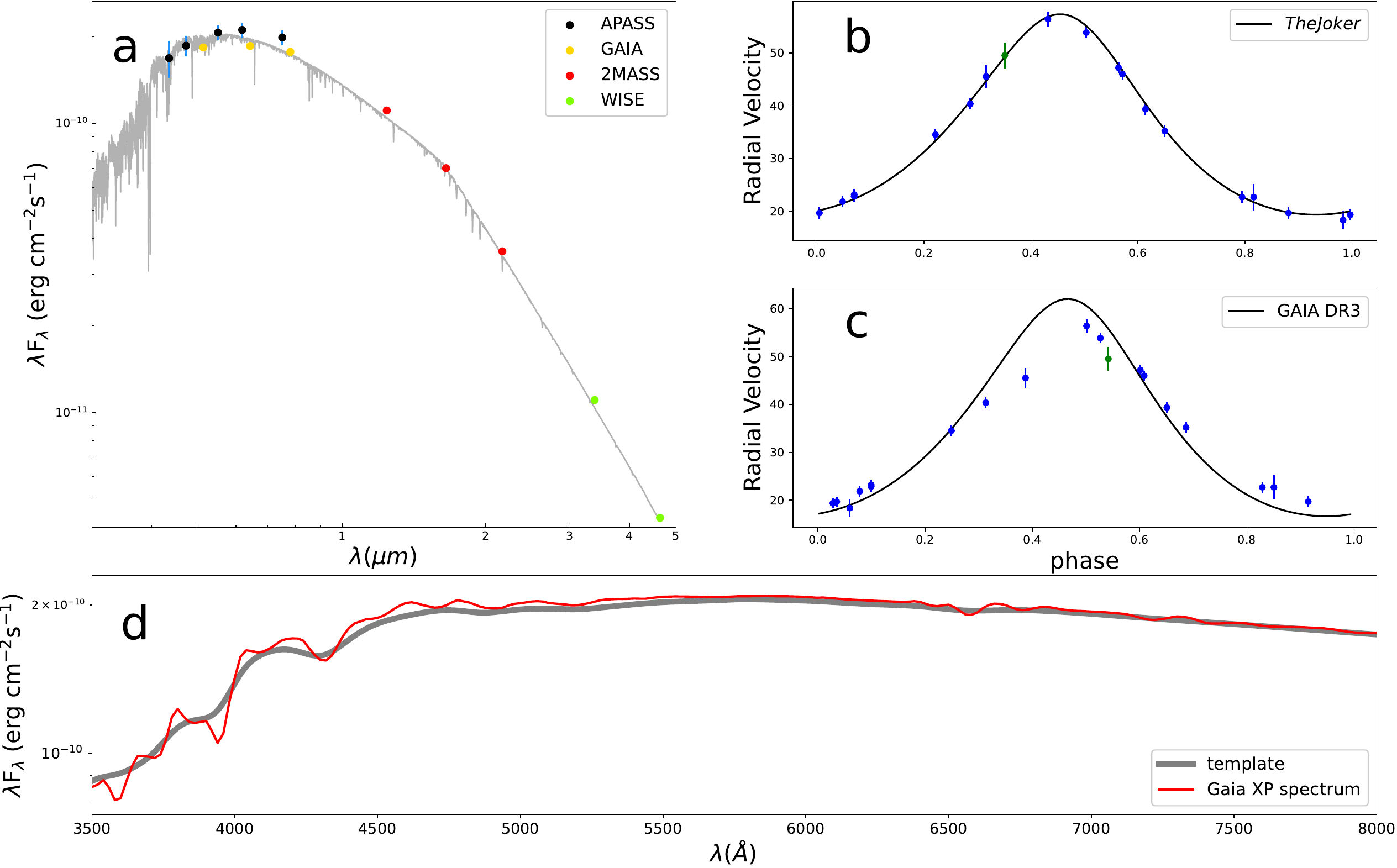}
    \caption{Panel a: SED fitting of G4031. Panel b: Folded LAMOST RV data and the RV curve from {\it The Joker}. The blue dots are the RV data from LAMOST MRS spectra, and the green dots are the RV data from LAMOST LRS spectra. Panel c: Folded LAMOST RV data and the RV curve from {\it Gaia} {\it $nss\_two\_body\_orbit$} solution. Panel d: Comparison of flux-calibrated {\it Gaia} XP spectrum and the Phoenix template ($T{\rm eff} = 6000 K$, ${\log}g = 4$ dex, [Fe/H]$=-0.5$).}
    \label{G4031.fig}
\end{figure*}

The stellar parameters can also be obtained by fitting the spectral energy distribution (SED). 
We used the Python module {\sc ARIADNE}\footnote{https://github.com/jvines/astroARIADNE} to estimate the stellar parameters of visible star, which can automatically fit the broadband photometry by using different stellar atmosphere models, such as Phoenix\footnote{ftp://phoenix.astro.physik.uni-goettingen.de/}, BTSettl\footnote{http://osubdd.ens-lyon.fr/phoenix/}, Castelli \& Kurucz\footnote{http://ssb.stsci.edu/cdbs/tarfiles/synphot3.tar.gz}, and Kurucz 1993\footnote{http://ssb.stsci.edu/cdbs/tarfiles/synphot4.tar.gz}.
We constructed the SED using magnitudes from various surveys: {\it Gaia} DR3 ($G$, $G_{\rm BP}$, and $G_{\rm RP}$), 2MASS ($J$, $H$, and $K_{\rm S}$), APASS ($B$, $V$, $g$, $r$, and $i$) and WISE ($W$1 and $W$2). 
The {\it Gaia} DR3 parallax and extinction $A_V$ were also used as input priors.
From the SED fitting (Figure \ref{G4031.fig}), we obtained stellar parameters for a G-type star with an effective temperature of $5910^{+36}_{-29} K$, surface gravity of $4.05^{+0.01}_{-0.01}$ dex, metallicity of $-0.37^{+0.01}_{-0.01}$ (Table \ref{astroARIADNE.tab}), which are in good agreement with spectroscopic results.

\begin{figure*}
    \center
    \includegraphics[width=1\textwidth]{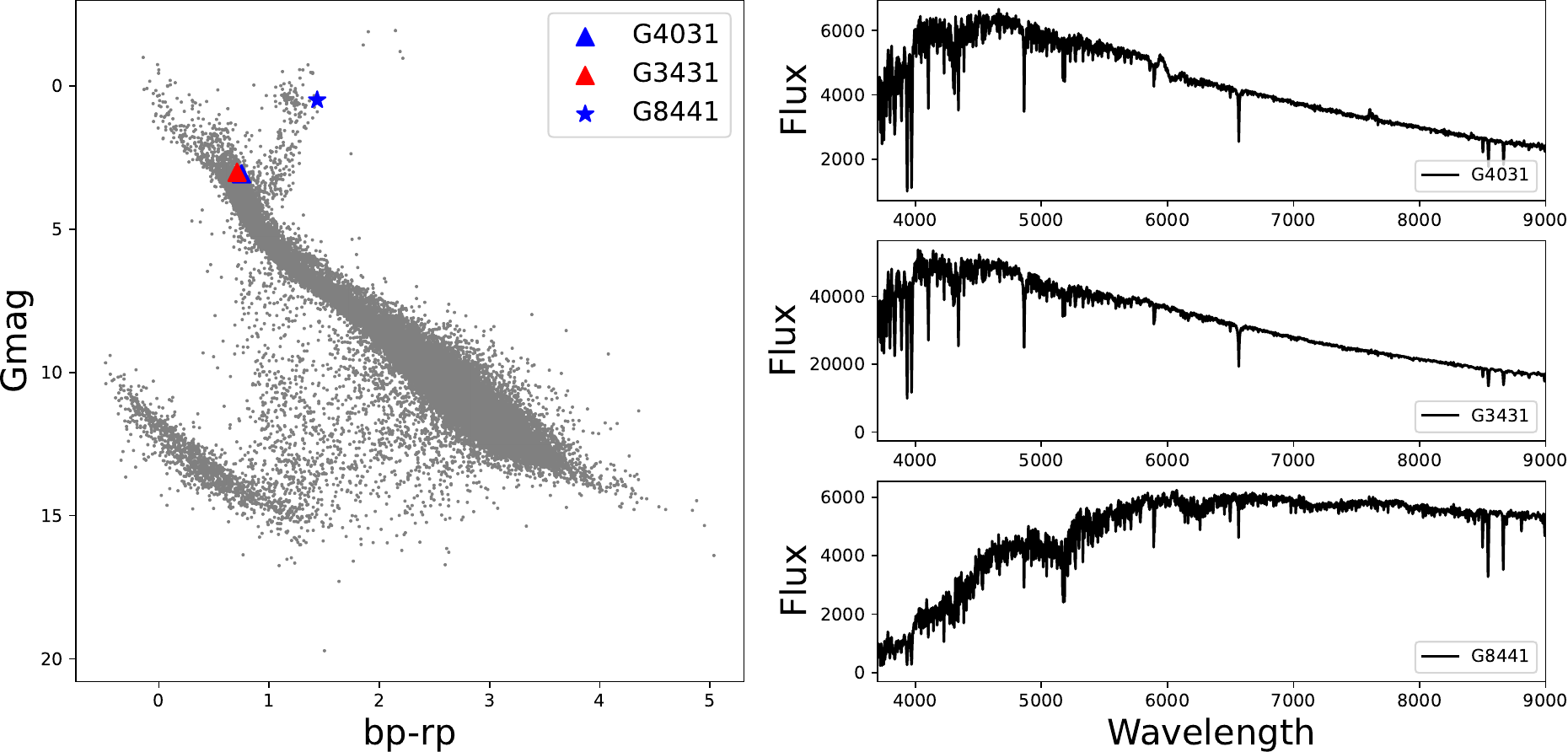}
    \caption{Left panel: Position of visible stars on the Hertzsprung–Russell diagram for G8441 (blue star), G3431 (red triangle) and G4031 (blue triangle). The gray points are from the {\it Gaia} DR3 with distance $d <$ 200 pc, $G_{\rm mag}$ between 4--20 mag, and galactic latitude $|b|$ $>$ 10. No extinction correction for those stars. 
    Right panel: LAMOST LRS observations of G4031, G3431 and G8441.}
    \label{HR_diagram.fig}
\end{figure*}

\begin{table}
\caption{Estimation of atmospheric parameters with different methods (LASP, DD-Payne, CNN) for our systems. The LASP results are from the LAMOST DR9 parameter catalog. \label{atmo_pars.tab}}
\centering
 \begin{tabular}{ccccc}
\hline\noalign{\smallskip}
name & Methods & $T_{\rm eff}$ ($K$) & log$g$ & [Fe/H] \\
\hline\noalign{\smallskip}
\multirow{4}*{G4031} & LASP (LRS)  & $5953{\pm 17}$ & $4.05{\pm 0.02}$ & $-0.37{\pm 0.01}$\\
& LASP (MRS)  & $6056{\pm 59}$ & $4.11{\pm 0.04}$ & $-0.40{\pm 0.05}$\\
& DD-Payne  & $5929{\pm 21}$ & $4.06{\pm 0.05}$ & $-0.43{\pm 0.04}$\\
& CNN  & $5915{\pm 99}$ & $4.12{\pm 0.09}$ & $-0.37{\pm 0.04}$\\
\hline
\multirow{4}*{G3431} & LASP (LRS) & $6402{\pm 13}$ & $4.23{\pm 0.02}$ & $-0.06{\pm 0.01}$\\
& LASP (MRS)  & $6419{\pm 43}$ & $4.24{\pm 0.02}$ & $-0.18{\pm 0.03}$\\
& DD-Payne  & $6352{\pm 10}$ & $4.27{\pm 0.04}$ & $-0.22{\pm 0.05}$\\
& CNN  & $6182{\pm 31}$ & $4.17{\pm 0.02}$ & $-0.23{\pm 0.02}$\\
\hline
\multirow{4}*{G8441} & LASP (LRS) & $4194{\pm 2}$ & $1.83{\pm 0.03}$ & $-0.74{\pm 0.01}$\\
& LASP (MRS)  & $4168{\pm 2}$ & $2.16{\pm 0.01}$ & $-0.75{\pm 0.02}$\\
& DD-Payne  & $4260{\pm 59}$ & $1.63{\pm 0.17}$ & $-0.67{\pm 0.09}$\\
& CNN  & $4233{\pm 55}$ & $1.86{\pm 0.11}$ & $-0.59{\pm 0.01}$\\
\noalign{\smallskip}\hline
\end{tabular}
\end{table}

\begin{table*}
\caption{Stellar parameters for our systems, including atmospheric parameters, distance, extinction and multi-band photometric magnitudes. \label{stellar_parameters.tab}}
\centering
\setlength{\tabcolsep}{4mm}
 \begin{tabular}{ccccc}
\hline\noalign{\smallskip}
Parameters & G4031 & G3431 & G8441 \\
\hline\noalign{\smallskip}
$T_{\rm eff}$ ($K$) & $5953{\pm 17}$ & $6402{\pm 13}$ & $4194{\pm 2}$ \\
log$g$ & $4.05{\pm 0.02}$ & $4.23{\pm 0.02}$ & $1.83{\pm 0.03}$ \\
${\rm [Fe/H]}$ & $-0.37{\pm 0.01}$ & $-0.06{\pm 0.01}$ & $-0.74{\pm 0.01}$ \\
\hline
$d$ (pc) & $720{\pm 10}$ & $338{\pm 3}$ & $857{\pm 12}$ \\
$\varpi$ (mas) & $1.3646{\pm 0.0183}$ & $2.9169{\pm 0.0346}$ & $1.1380{\pm 0.017}$ \\
$E(B-V)$ & 0.02 & 0.03 & 0.01 \\
\hline
$G$ (mag) & $12.35{\pm 0.001}$ & $10.73{\pm 0.003}$ & $10.18{\pm 0.003}$ \\
$BP$ (mag)  & $12.65{\pm 0.001}$ & $11.02{\pm 0.003}$ & $10.86{\pm 0.009}$ \\
$RP$ (mag)  & $11.90{\pm 0.001}$ & $10.27{\pm 0.004}$ & $9.41{\pm 0.008}$ \\
$J$ (mag)  & $11.35{\pm 0.018}$ & $9.74{\pm 0.022}$ & $8.28{\pm 0.023}$ \\
$H$ (mag)  & $11.07{\pm 0.022}$ & $9.46{\pm 0.022}$ & $7.56{\pm 0.061}$ \\
$K_{\rm S}$ (mag)  & $11.02{\pm 0.019}$ & $9.40{\pm 0.020}$ & $7.38{\pm 0.034}$ \\
\noalign{\smallskip}\hline
\end{tabular}
\end{table*}

\begin{table}
\caption{Estimation of stellar parameters from SED fitting for our systems. \label{astroARIADNE.tab}}
\centering
\setlength{\tabcolsep}{1.pt}
 \begin{tabular}{cccc}
\hline\noalign{\smallskip}
parameters & G4031 & G3431 & G8441 \\
\hline\noalign{\smallskip}
$T_{\rm eff}$ ($K$) & $5910^{+36}_{-29}$ & $6156^{+69}_{-39}$ & $4144^{+23}_{-20}$ \\
log$g$ & $4.05{\pm 0.01}$ & $4.22^{+0.04}_{-0.03}$ & $1.86^{+0.11}_{-0.09}$ \\
${\rm [Fe/H]}$ & $-0.37{\pm 0.01}$ & $-0.06{\pm 0.01}$ & $-0.77{\pm 0.10}$ \\
$D$ (pc) & $721^{+11}_{-13}$ & $338^{+6}_{-7}$ & $864^{+40}_{-48}$ \\
$R$ ($R_{\odot}$) & $2.03{\pm 0.04}$ & $1.93^{+0.06}_{-0.05}$ & $16.55^{+0.79}_{-0.96}$ \\
\noalign{\smallskip}\hline
\end{tabular}
\end{table}

Furthermore, we calculated stellar mass using two methods: one method involved the use of stellar evolution models, referred to as evolutionary mass; the other method utilized observed photometric and spectroscopic parameters, referred to as gravitational mass.

First, we used the {\it isochrones}, a Python module \citep{2015ascl.soft03010M}, to calculate evolutionary mass of G4031 by fitting the spectroscopic and photometric data. 
We used the atmospheric parameters ($T_{\rm eff}$, log$g$, [Fe/H]), {\it Gaia} parallax \citep{2018AA...616A...1G}, multi-band magnitudes ($G$, $G_{\rm BP}$, $G_{\rm RP}$, $J$, $H$, and $K_{\rm S}$), and extinction $A_V$ ($= 3.1 \times E(B-V)$) as the input parameters. 
The evolutionary mass and radius of visible star from {\it isochrones} are $1.46^{+0.02}_{-0.02} M_{\odot}$ and $2.04^{+0.03}_{-0.03} R_{\odot}$, respectively.

Second, we used {\it Gaia} and {\it 2MASS} magnitudes ($G$, $G_{\rm BP}$, $G_{\rm RP}$, $J$, $H$, and $K_{\rm S}$) and the effective temperature to calculate the gravitational mass following $M = gR^2/G$ \citep[See details in ][]{2022ApJ...938...78L}.
The final mass and corresponding uncertainty were calculated as the average values and standard deviations of the masses derived from different bands.
%
The gravitational mass of G4031 is $1.62{\pm 0.02} M_{\odot}$.

\subsection{Kepler orbit}

We averaged the Barycentric Julian Dates (BJDs) and RVs taken on the same day (Appendix \ref{rvdata_appendix.sec}) for the purpose of RV fitting.
{\it The Joker} \citep{2017ApJ...837...20P}, a custom Markov chain Monte Carlo sampler, was employed to fit the Keplerian orbit. 
Figure \ref{G4031.fig} displays the phase-folded RV data along with the best-fit RV curve.
Table \ref{keplerorbit.tab} presents the orbital parameters of G4031 obtained through {\it The Joker} fitting, including orbital period $P$, eccentricity $e$, argument of periastron $\omega$, mean anomaly at the first exposure $M_{\rm0}$, RV semi-amplitude $K$, and systematic RV $\nu_{\rm 0}$.
The fitting results differ slightly from those obtained from the table {\it $nss\_two\_body\_orbit$} of {\it Gaia} DR3.

The binary mass function can be calculated following
\begin{equation}
    f(M) = \frac{M_{2} \, \textrm{sin}^3 i} {(1+q)^{2}} = \frac{P \, K_{1}^{3} \, (1-e^2)^{3/2}}{2\pi G},
\label{mass_function.equ}
\end{equation}
\noindent
where $M_{2}$ is the mass of the unseen star, $q = M_{1}/M_{2}$ is the mass ratio of this system, and $i$ is the systematic inclination angle. 
We obtained a mass function of $\approx$0.09 $M_{\odot}$.
By using the evolutionary mass of $1.46^{+0.02}_{-0.02} M_{\odot}$ and gravitational mass of $1.62{\pm 0.02} M_{\odot}$ of the visible star, we determined the minimum mass of the unseen object as $0.77\pm0.03 M_{\odot}$ and $0.82\pm0.02 M_{\odot}$, respectively.
Using the orbital parameters from {\it Gaia} DR3, the mass function is $\approx$0.16 $M_{\odot}$ while the minimum mass of the unseen star would be $0.98\pm0.09 M_{\odot}$ or $1.04\pm0.09 M_{\odot}$.
These results suggest G4031 may contain a white dwarf.

\begin{figure}
    \center
    \includegraphics[width=0.49\textwidth]{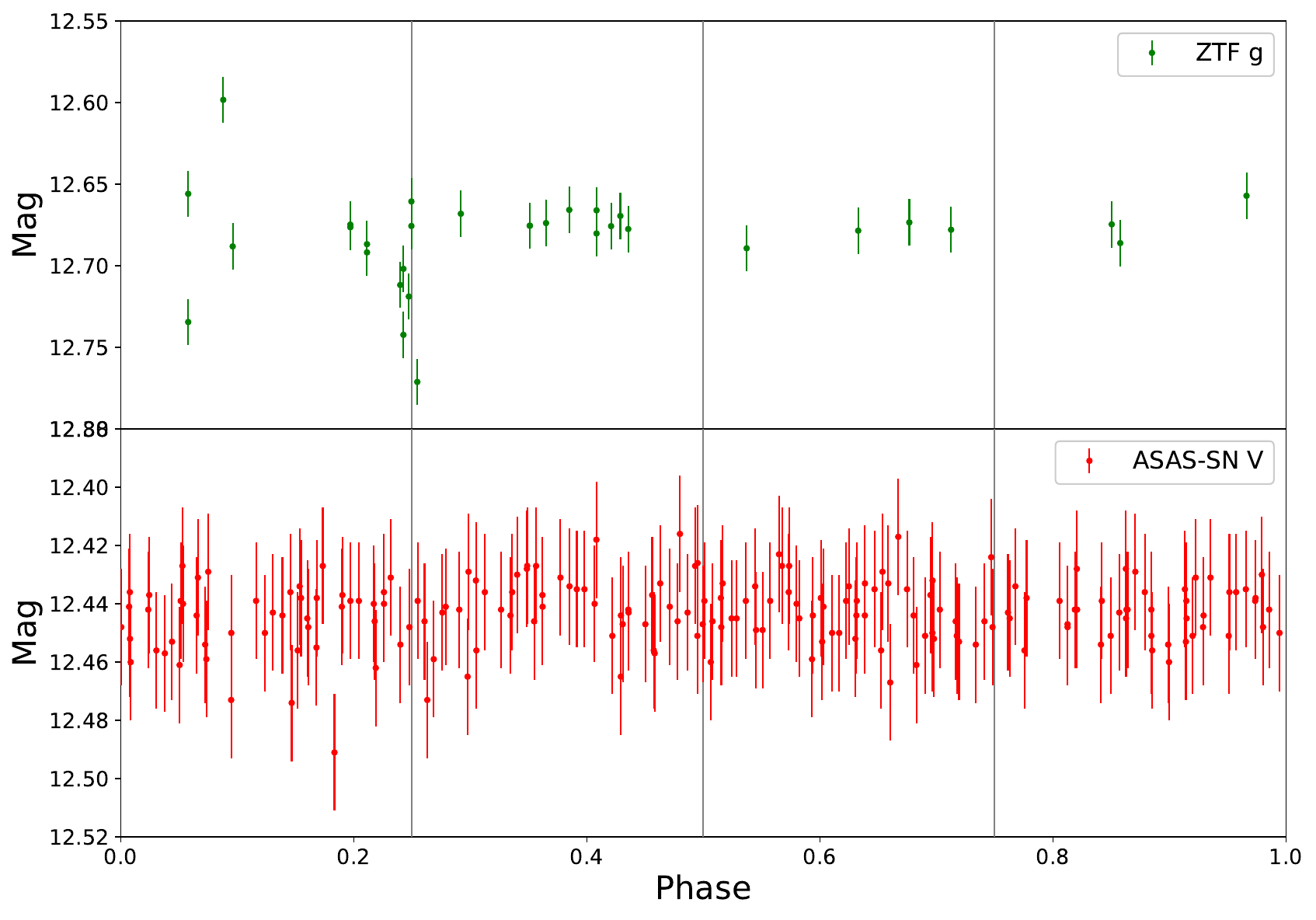}
    \caption{Folded ASAS-SN $V$-band and ZTF $g$-band light curves of G4031 with the orbital period of $\approx$137.98 day.}
    \label{4031_lcs.fig}
\end{figure}

G4031 has been observed by Zwicky Transient Facility \citep[ZTF;][]{2019PASP..131a8002B} in $g$ band and All-Sky Automated Survey for Supernovae \citep[ASAS-SN;][]{2017PASP..129j4502K} in $V$ band. 
No clear variation caused by ellipsoidal deformation can be seen in the folded light curves (Figure \ref{4031_lcs.fig}).
Therefore, no further information such as the orbital inclination can be constrained by the light curve.

\begin{table*}
\caption{Keplerian orbit fitting results from the {\it The Joker} and {\it Gaia} DR3 table {\it $nss\_two\_body\_orbit$}. The minimum mass of secondary ($M_{\rm 2,min}$) is calculated by using the evolutionary mass of the visible star. \label{keplerorbit.tab}}
\centering
\setlength{\tabcolsep}{1.5mm}
 \begin{tabular}{ccccccccccc}
\hline\noalign{\smallskip}
name & Methods & $P$ (day) & $e$ & $\omega$ & $M{\rm 0}$ & $K_{\rm 1}$ (km/s) & $\nu{\rm 0}$ (km/s) & $f$(m) & $M_{\rm 2,min} (M_{\odot})$\\
\hline\noalign{\smallskip}
 \multirow{2}*{G4031} & {\it The Joker} & $137.98^{+0.66}_{-0.61}$ & $0.215^{+0.021}_{-0.021}$ & $0.12^{+0.08}_{-0.07}$ & $3.00^{+0.09}_{-0.10}$ & $19.00^{+0.40}_{-0.41}$ & $34.22^{+0.27}_{-0.28}$ & $0.092{\pm 0.006}$ & $0.77{\pm 0.03}$\\
&  {\it Gaia} DR3 & $140.08{\pm 0.35}$ & $0.227{\pm 0.047}$ & - & - & $22.71{\pm 1.42}$ & $35.67{\pm 0.90}$ & $0.158{\pm 0.031}$ & $0.98{\pm 0.09}$\\
\hline
 \multirow{2}*{G3431} & {\it The Joker} & $120.56^{+0.20}_{-0.19}$ & $0.130^{+0.030}_{-0.020}$ & $2.89^{+0.30}_{-0.26}$ & $-2.39^{+0.36}_{-0.32}$ & $29.83^{+1.05}_{-0.97}$ & $64.12^{+0.63}_{-0.66}$ & $0.324{\pm 0.038}$ & $1.36{\pm 0.09}$\\
&  {\it Gaia} DR3 & $120.87{\pm 0.10}$ & $0.193{\pm 0.017}$ & - & - & $32.11{\pm 0.60}$ & $63.74{\pm 0.41}$ & $0.392{\pm 0.023}$ & $1.49{\pm 0.05}$\\
\hline
 \multirow{2}*{G8441} & {\it The Joker} & $46.81^{+0.12}_{-0.08}$ & $0.020^{+0.030}_{-0.020}$ & $0.46^{+1.44}_{-1.59}$ & $0.72^{+1.36}_{-1.85}$ & $55.71^{+2.44}_{-2.58}$ & $17.99^{+1.51}_{-1.40}$ & $0.838{\pm 0.116}$ & $1.25{\pm 0.17}$\\
 & {\it Gaia} DR3 & $46.88{\pm 0.01}$ & $0.010{\pm 0.003}$ & - & - & $51.70{\pm 0.15}$ & $20.50{\pm 0.10}$ & $0.671{\pm 0.006}$ & $1.07{\pm 0.10}$\\
\noalign{\smallskip}\hline
\end{tabular}
\end{table*}

\subsection{The nature of the unseen object}

We calculated the size of Roche-lobe by using the standard Roche-lobe approximation \citep{1983ApJ...268..368E},
\begin{equation}
\frac{R_1}{a} =  
    \frac{0.49 q^{2/3}}{0.6q^{2/3} + \ln(1 + q^{1/3})},
\label{eq.equ} 
\end{equation} 
where $a = (1+q) a_1 = (1+q) P (1-e^2)^{1/2} K_1/(2 \pi {\rm \sin}i)$ is the separation of binary system. 
The filling factor is $\approx 3.2\%$ assuming $i = 90^{\circ}$, indicating this system is a detached binary system.

For the visible star, the absolute magnitude in the $G$ band is $G=3.04$ mag; while for a main-sequence star with a mass of $0.77 M_{\odot}$ or $0.98 M_{\odot}$, the absolute magnitude in the $G$ band would be around 6.20 mag or 4.80 mag. 
By comparing the flux-calibrated {\it Gaia} XP spectrum (Huang et al. 2023, in preparation) with the Phoenix template, no flux excess is observed (Figure \ref{G4031.fig}), indicating the companion of G4031 is a compact object if the secondary has a mass of $0.98 M_{\odot}$.
However, if the secondary has a mass of $0.77 M_{\odot}$, it cannot be distinguished based on the SED or spectral continuum.
Given this uncertainty, we decided to investigate the binary nature of G4031 with the LAMOST MRS observations by using the spectral disentangling method.

We employed the algorithm of spectral disentangling proposed by \citet{1994A&A...281..286S}. 
Before applying this algorithm, we performed preliminary tests to determine the detection limits for our targets. 
We derived the effective temperature and surface gravity of the secondary (assuming a main-sequence star) using the minimum mass\footnote{http://www.pas.rochester.edu/\~emamajek/EEM\_dwarf\_UBV-IJHK\_colors\_Teff.txt} following different mass ratio (Table \ref{test_dis.tab}). 
The corresponding Phoenix\footnote{https://phoenix.astro.physik.uni-goettingen.de} model was used as the theoretical spectra of the secondary. 
Then, we combined the observed spectra and theoretical spectra, with a reduced resolution of $R=$ 7500 and different rotational broadening ($v{\rm sin}i =$ 10 km/s, 50 km/s, 100 km/s and 150 km/s), to produce synthetic binary spectra.
For the simulated binary spectra, the wavelength range of 6400 \AA \ to 6600 \AA \ was used for spectral disentangling. 
The results by visual check are presented in Table \ref{test_dis.tab}. 
It can be seen that for G4031, when $v{\rm sin}i$ is less than 100 km/s, this method can successfully separate the spectra of primary and secondary components.

\begin{figure*}
    \center
    \includegraphics[width=1\textwidth]{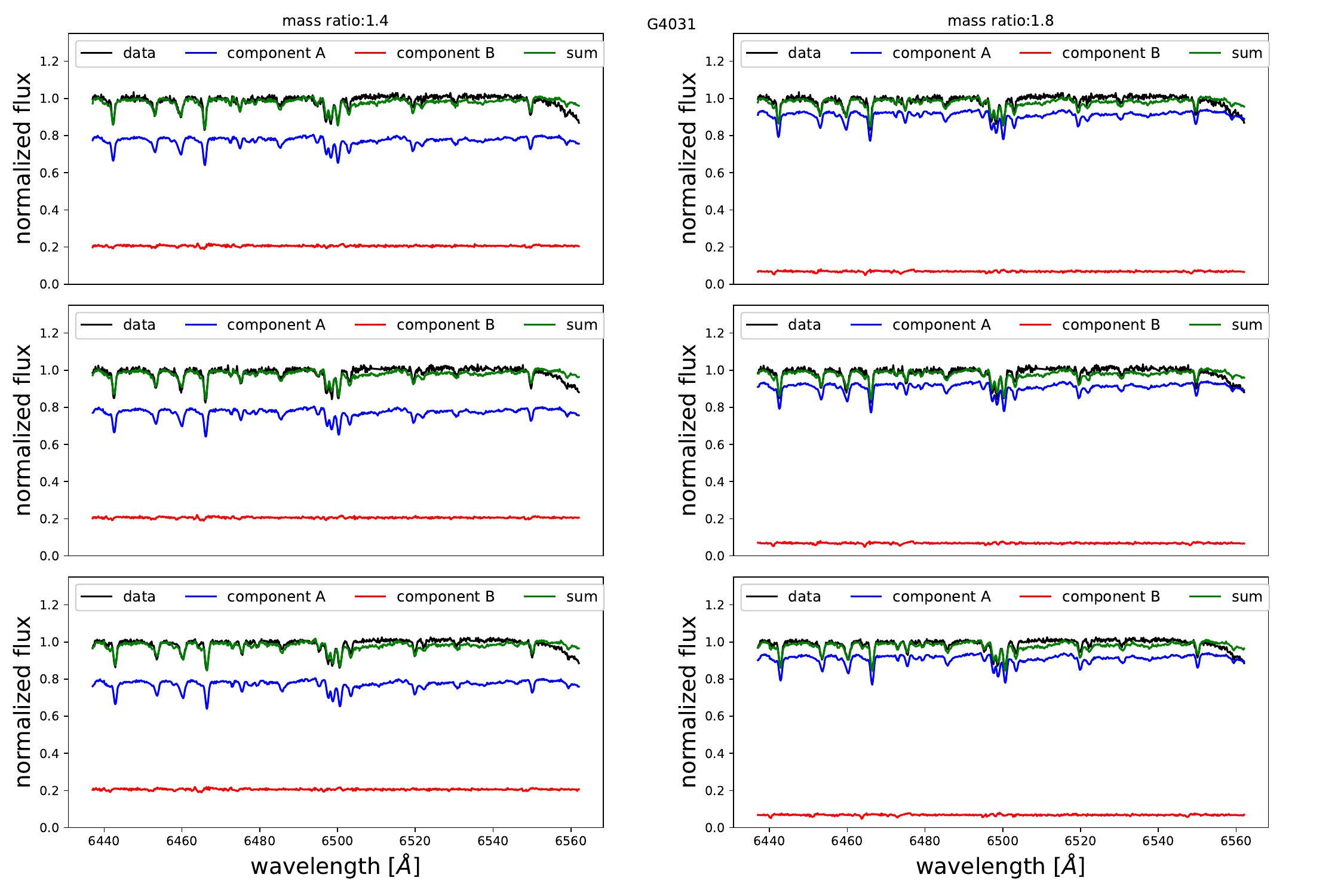}
    \caption{Left panel: spectral disentangling of G4031 with $q =$ 1.4. Right Panel: spectral disentangling with $q =$ 1.8. The vertical panels show spectra in different phases (close to the minimum or maximum RV phase of the visible star). The blue lines mark the reconstructed spectra of the visible star, while the red lines represent the second component in each spectra. The green lines are the sum of the two components, and the black lines represent the observed spectra.}
    \label{dis_4031.fig}
\end{figure*}

The MRS observations within the wavelength range of 6400 Å to 6600 Å were utilized for spectral disentangling.
As an example, Figure \ref{dis_4031.fig} shows the results for a mass ratio of 1.4 and 1.8.
No feature of absorption spectrum is apparent for an additional component (red lines in Figure \ref{dis_4031.fig}), implying that the component star is a compact object.

We also used single- and binary-star spectral models (Liu et al. 2023, in preparation) to perform fitting for the LRS observations. 
In brief, the spectra from LAMOST LRS and the stellar parameters from the Apache Point Observatory Galactic Evolution Experiment (APOGEE) DR16 were used as the training set to develop a model using the neural network version of Stellar LAbel Machine (SLAM) \citep{2020ApJS..246....9Z,2020RAA....20...51Z} for single-star spectra.
The binary-star model was then created by combining two single-star models with appropriate radial velocities.
Spectral observations were fitted with both the single- and binary-star models, and their corresponding $\chi^{2}$ values were compared to determine the best fit.
The results of the semi-empirical spectroscopic experiments show that one system is a single star if the difference in $\chi^{2}$ (i.e., $\Delta \chi^{2}$) between the single-star and binary-star fitting is less than 0.4.
For G4031, the $\Delta \chi^{2}$ value is about $-$0.04, further evidencing that it includes one compact object.

\begin{table}
\caption{Spectral disentangling test results by visual check.  "\ding{52}\ding{52}" means the spectra are well disentangled, "\ding{52}" means the spectra can be disentangled but in low significance, and "\ding{53}" means the spectra can not be disentangled. \label{test_dis.tab}}
\centering
\setlength{\tabcolsep}{4pt}
 \begin{tabular}{cccc}
\hline\noalign{\smallskip}
Object & $q$ & $v{\rm sin}i$ & 
\begin{tabular}{c}Disentangling  \\ results \end{tabular} \\
\hline\noalign{\smallskip}
  \multirow{12}*{G4031} & \multirow{4}*{1.4} & 10 & \ding{52}\ding{52}\\
 & & 50 & \ding{52}\ding{52}\\
 & & 100 & \ding{52}\\
  & & 150 & \ding{52}\\
   \cline{2-4}
&  \multirow{4}*{1.6}  & 10 & \ding{52}\ding{52}\\
& & 50 & \ding{52}\\
& & 100 & \ding{52}\\
 & & 150 & \ding{53}\\
  \cline{2-4}
&  \multirow{4}*{1.8}  & 10 & \ding{52}\\
& & 50 & \ding{52}\\
& & 100 & \ding{53}\\
 & & 150 & \ding{53}\\
 \hline
  \multirow{12}*{G3431} & \multirow{4}*{1} & 10 & \ding{52}\ding{52}\\
 & & 50 & \ding{52}\ding{52}\\
 & & 100 & \ding{52}\ding{52}\\
  & & 150 & \ding{52}\ding{52}\\
   \cline{2-4}
&  \multirow{4}*{1.1}  & 10 & \ding{52}\ding{52}\\
& & 50 & \ding{52}\ding{52}\\
& & 100 & \ding{52}\ding{52}\\
 & & 150 & \ding{52}\ding{52}\\
  \cline{2-4}
&  \multirow{4}*{1.3}  & 10 & \ding{52}\ding{52}\\
& & 50 & \ding{52}\ding{52}\\
& & 100 & \ding{52}\\
 & & 150 & \ding{52}\\
\noalign{\smallskip}\hline
\end{tabular}
\end{table}

\section{G3431}
\label{g3431.sec}

\subsection{Stellar parameters of the visible star}

\begin{figure*}
    \center
    \includegraphics[width=1\textwidth]{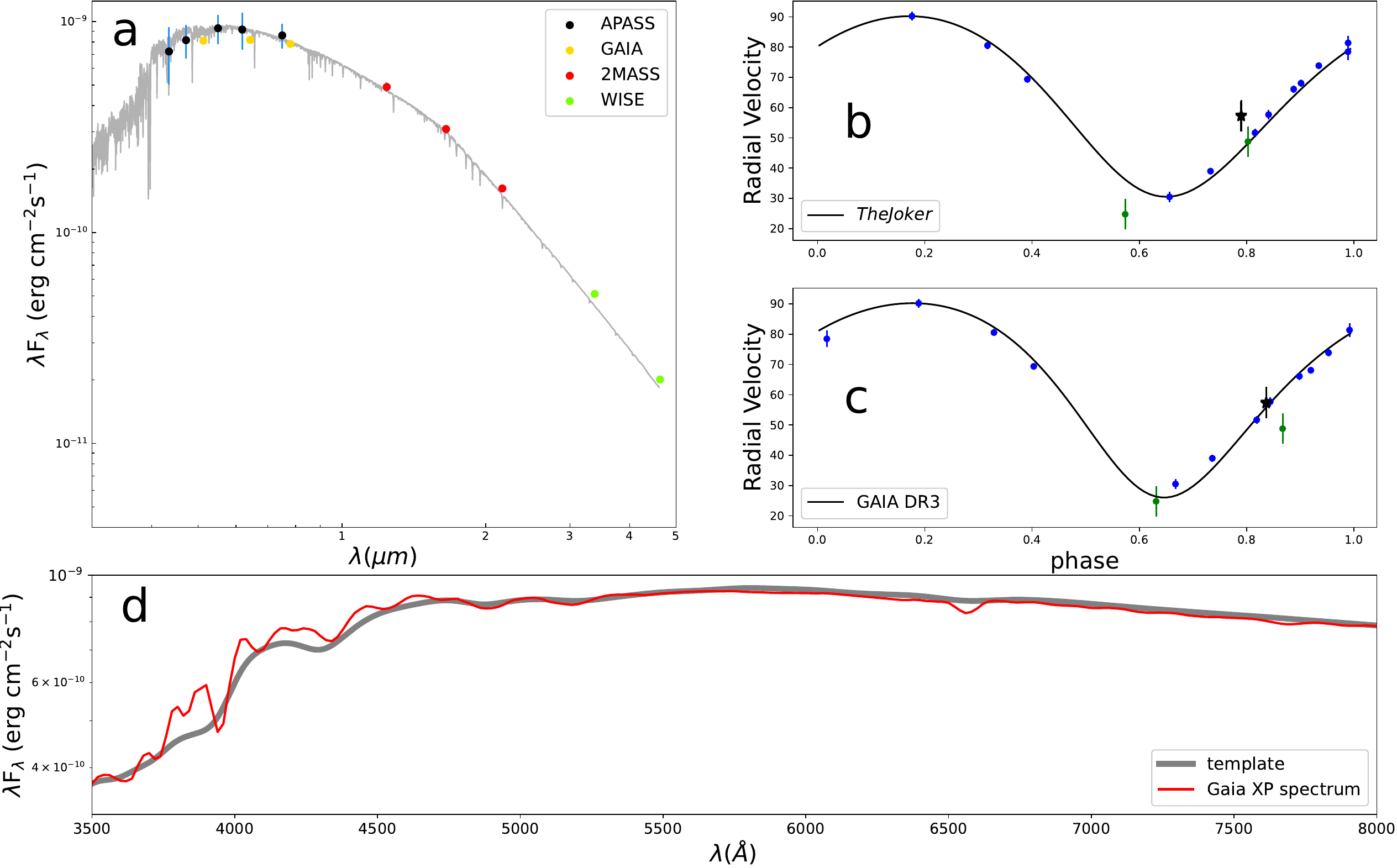}
    \caption{Panel a: SED fitting of G3431. Panel b: Folded LAMOST RV data and best-fit RV curve from {\it The Joker}. The blue and green dots are the RV data from LAMOST MRS and LRS observations, and the black star marks the RV data from 2.16 m observation. Panel c: Folded RV data and the RV curve from {\it Gaia} {\it $nss\_two\_body\_orbit$} solution. Panel d: Comparison of the flux-calibrated {\it Gaia} XP spectrum and the Phoenix template spectrum  ($T_{\rm eff} = 6200 K$, log$g = 4.5$ dex, [Fe/H]=0.0) for G3431.}
    \label{G3431.fig}
\end{figure*}

The visible star of G3431 is a G-type main-sequence star. 
The stellar parameters from the LASP (LRS) method are used in the following analysis, with an effective temperature of $T_{\rm eff} = 6402{\pm 13} K$, surface gravity of log$g$ $=4.23{\pm 0.02}$ dex, and metallicity of [Fe/H] $=-0.06{\pm 0.01}$ (Table \ref{stellar_parameters.tab}). 
According to {\it Gaia} DR3, the estimated distance to G3431 is approximately 338 pc.
The Pan-STARRS DR1 3D dust map indicates an extinction with $E(B-V) =0.03$ at this distance.
SED fitting returns stellar parameter estimations for the visible star,  with $T_{\rm eff} = 6156^{+69}_{-39} K$, log$g$ $=4.22^{+0.04}_{-0.03}$ dex, [Fe/H] $=-0.06^{+0.01}_{-0.01}$ (Table \ref{astroARIADNE.tab} and Figure \ref{G3431.fig}).

The {\it isochrones} fitting based on evolutionary models yields an estimated evolutionary mass and radius for G3431 of $1.42^{+0.03}_{-0.03} M_{\odot}$ and $1.83^{+0.03}_{-0.03} R_{\odot}$, respectively.
On the other hand, the gravitational mass of G3431, obtained by averaging six gravitational mass estimations, is determined to be $2.06{\pm 0.13} M_{\odot}$, which differs from the evolutionary mass estimation.
It is important to note that even within the gravitational mass measurements, there are discrepancies between the results derived from {\it Gaia} bands ($\approx1.88\ M_{\odot}$) and those from 2MASS bands ($\approx2.24\ M_{\odot}$).
By reviewing the calculation process, we found that the bolometric magnitudes from {\it Gaia} bands ($BP$, $G$, $RP$) are 3.13, 3.09, 3.04, while the bolometric magnitudes from 2MASS bands ($J$, $H$, $K_{\rm S}$) are 2.93, 2.89, 2.86. 
This gradual variation with wavelength can be attributed to the measurement accuracy of stellar parameters, such as $T_{\rm eff}$ and $E(B-V)$.
We noticed that the difference in temperature between different methods is about 250 K (Table \ref{atmo_pars.tab}), and the temperature from SED fitting is lower than spectral results.
In this case, the mass measurement from stellar model fitting is more reliable than the mass calculated with atmospheric parameters.
Additionally, the evolutionary mass of $\approx$1.42 $M_{\odot}$ is more reasonable for a late-F main-sequence star.
Taking all these factors into consideration, we adopted the evolutionary mass measurement for further analysis.

\subsection{Kepler orbit}

\begin{figure}
    \center
    \includegraphics[width=0.49\textwidth]{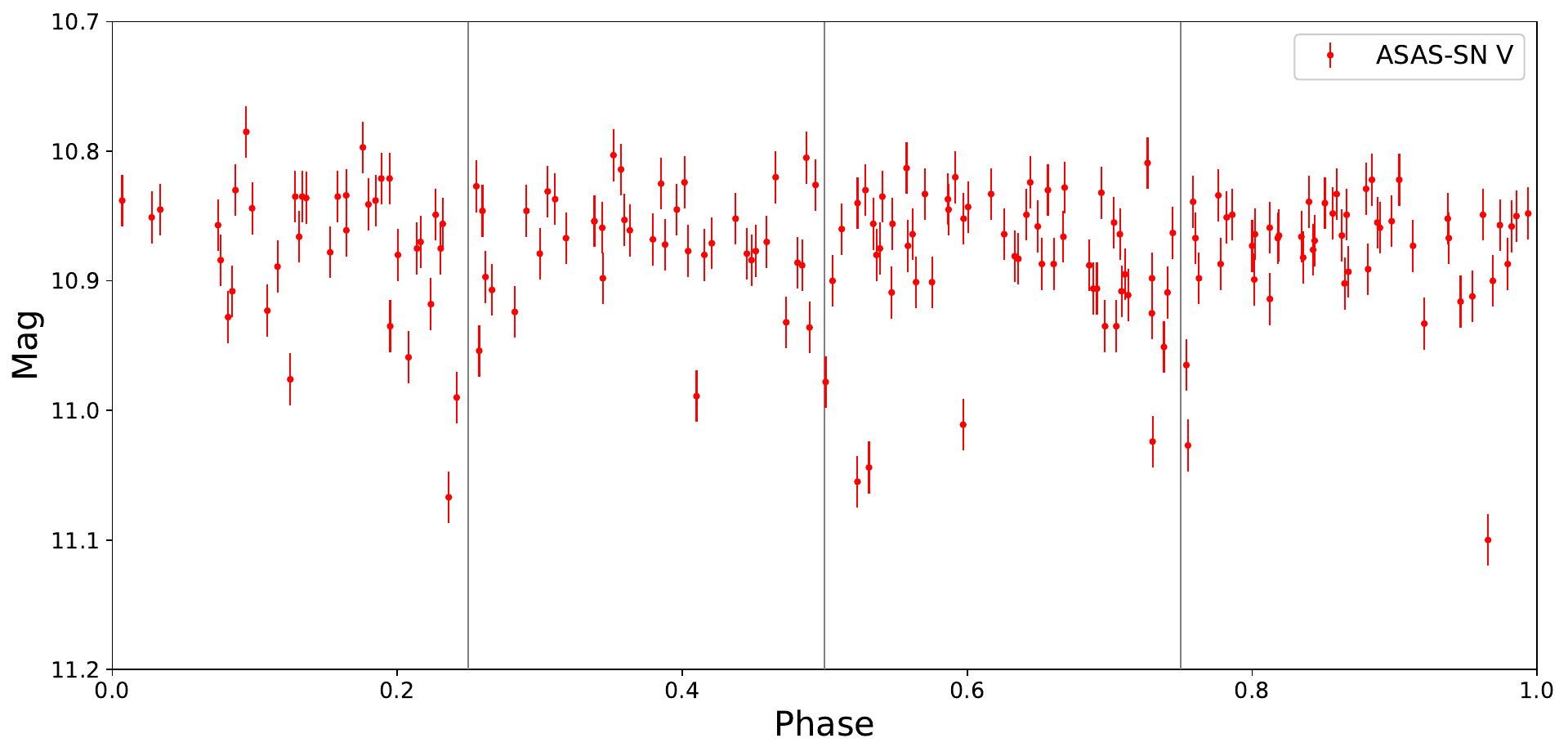}
    \caption{Folded ASAS-SN $V$-band light curve of G3431 by using the orbital period of $\approx$120.56 day.}
    \label{3431_lcs.fig}
\end{figure}

The Barycentric Julian Dates (BJDs) and RV values taken on the same day (Appendix \ref{rvdata_appendix.sec}) were averaged to improve the accuracy of the RV fitting.
The phase-folded RV data and the best-fit RV curve from {\it The Joker} are shown in Figure \ref{G3431.fig}.
Based on the RV fitting results, we obtained a mass function of $\approx$0.32 $M_{\odot}$ and a minimum mass of $1.36\pm0.09 M_{\odot}$ for the unseen star, which are slightly lower than the values calculated from {\it Gaia} DR3 solution.
As G4031, no clear variation can be seen in the light curve of G3431 (Figure \ref{3431_lcs.fig}).

\subsection{The nature of the unseen object}

Using the Roche-lobe approximation (Eq. \ref{eq.equ}), we estimated that the filling factor of G3431 is approximately $3.4\%$ (assuming $i = 90^{\circ}$), suggesting this system is a detached binary system.
The possibility that G3431 contains two main-sequence stars can be ruled out according to the luminosity ratio. 
For instance, the absolute $G$-band magnitude of the visible star is $G =$ 3.06 mag, while the absolute magnitude of a main-sequence star with a mass of $1.36 M_{\odot}$ is $G =$ 3.10 mag.

\begin{figure*}
    \center
    \includegraphics[width=1\textwidth]{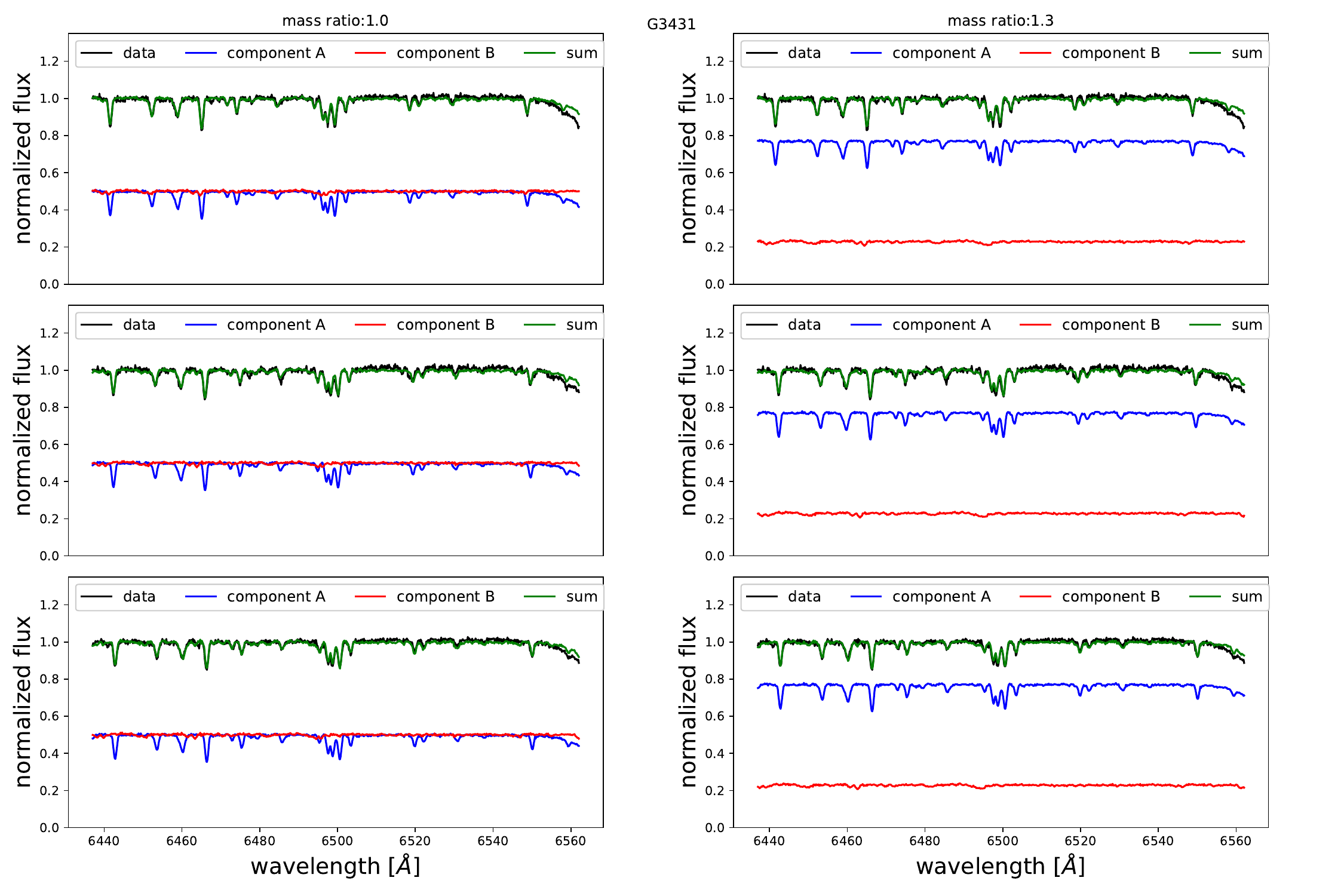}
    \caption{Left panel: spectral disentangling of G3431 with $q =$ 1.0. Right panel: spectral disentangling with $q =$ 1.3. Symbols are as in Figure \ref{dis_4031.fig}.}
    \label{dis_3431.fig}
\end{figure*}

We also applied the spectral disentangling technique to the observed spectra of G3431.
The results of the spectral disentangling tests (Table \ref{test_dis.tab}) indicate that for binary systems with two normal stars having a mass set similar to G3431, this method can successfully separate the binary components including two normal stars with high significance when $v{\rm sin}i$ is less than 150 km/s.
No additional component with an optical absorption spectrum can be detected (Figure \ref{dis_3431.fig}), excluding the possibility of a normal binary system.
Furthermore, the $\Delta \chi^{2}$ values of G3431 from single-star and binary fitting are about $-$0.18 and $-$0.24 for its two LRS observations, indicating that these spectra are more likely from a single star.

\section{G8441}
\label{g8441.sec}

\subsection{Stellar parameters of the visible star}
\label{8441_info.sec}

\begin{figure*}
    \center
    \includegraphics[width=1\textwidth]{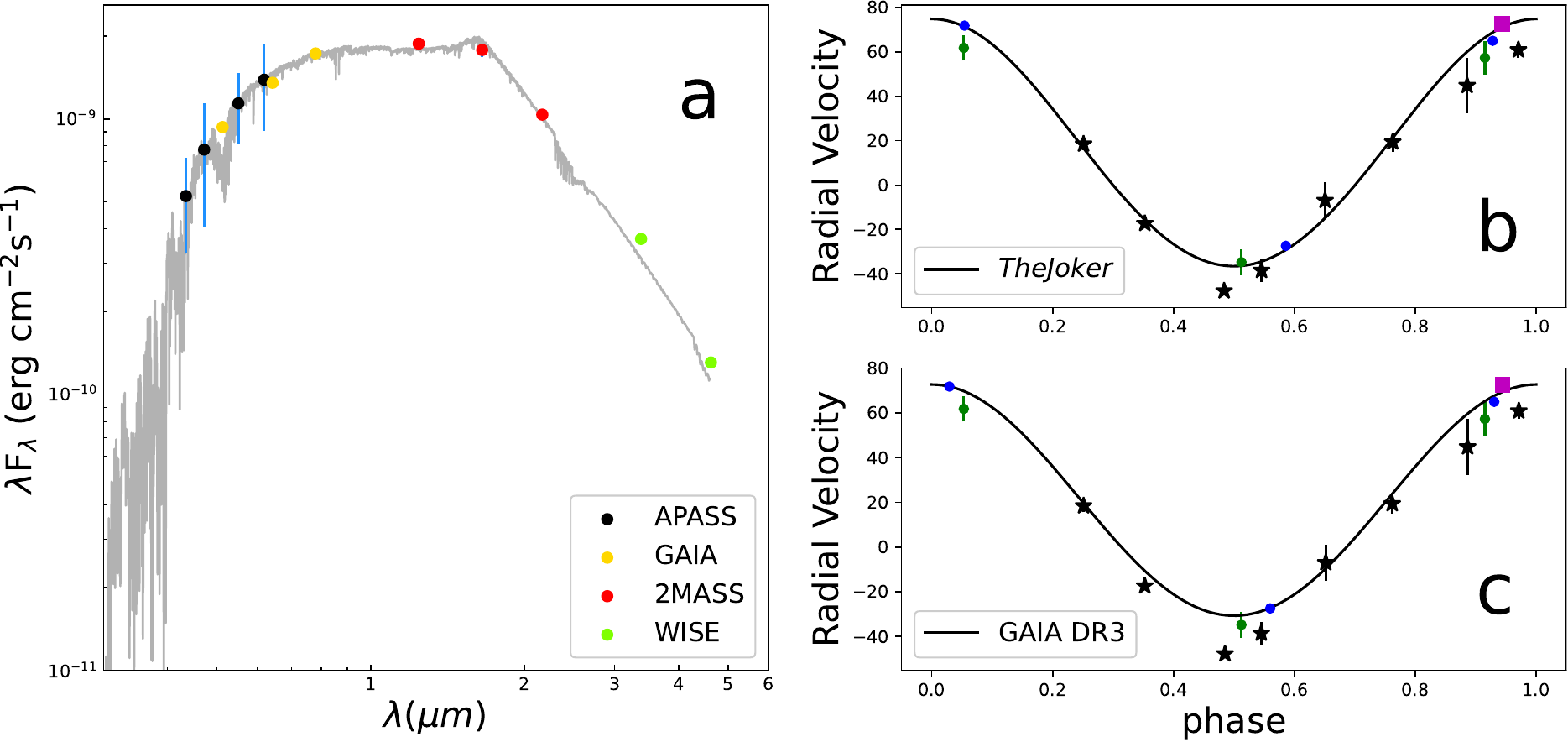}
    \caption{Panel a: SED fitting of G8441. Panel b: Folded LAMOST RV data and the RV curve from {\it Gaia} {\it $nss\_two\_body\_orbit$} solution. The green and blue dots are the RV data from LAMOST LRS and MRS, respectively. The purple square marks the RV data from P200, while the black stars represent the RV data from 2.16 m telescope.}
    \label{G8441.fig}
\end{figure*}

The stellar parameters of G8441 obtained from LASP (LRS) are as follows: $T_{\rm eff} = 4194{\pm 2}\ K$, log$g$ $=1.83{\pm 0.03}$ dex, and [Fe/H] $=-0.74{\pm 0.01}$ (Table \ref{stellar_parameters.tab}).
According to {\it Gaia} DR3, the estimated distance to G8441 is approximately 857 pc, and the Pan-STARRS DR1 3D dust map suggests an extinction of $E(B-V) \approx 0.01$ at this distance.
SED fitting using the {\sc ARIADNE} package (Figure \ref{G8441.fig}a) yields an effective temperature of $4144^{+23}_{-20}\ K$, surface gravity of $1.86^{+0.11}_{-0.09}$ dex, and metallicity of $-0.77^{+0.10}_{-0.10}$, which are in good agreement with the spectroscopic estimation.

Unlike G4031 and G3431, the light curves of G8441 exhibit significant ellipsoidal variations (Section \ref{G8441orbit}).
The large amplitude of variation in $V$-band light curve ($\gtrsim$ 0.4 mag) implies that the visible star of G8441 is (close to) Roche lobe-filling.
As a result, the single-star evolution model is unsuitable for assessing the mass of the visible star.
For a stripped star with Roche lobe overflow, its density can be calculated following \citep{2002apa..book.....F},
\begin{equation}
\bar{\rho} = 110P^{-2}_{\rm hr}\ {\rm g}\ {\rm cm^{-3}}.
\end{equation}
Using the radius from SED fitting ($R =$ 16.55 $R_{\odot}$; Table \ref{astroARIADNE.tab}), we calculated the mass of the visible star to be $M = 4 \pi \bar{\rho} R^{3} /3 \approx 0.28{\pm 0.03}\ M_{\odot}$.

\subsection{Kepler orbit}
\label{G8441orbit}

The phase-folded RV data and the best-fit RV curve from {\it The Joker} are shown in Figure \ref{G8441.fig}.
The RV fitting returns a mass function of $\approx$0.84 $M_{\odot}$ and a minimum mass of $1.25\pm0.09\ M_{\odot}$ for the unseen star, while using the orbital parameters of {\it Gaia} DR3 (Table \ref{keplerorbit.tab}), the mass function is $\approx$0.67 $M_{\odot}$ and the minimum mass of the unseen star is $1.07\pm0.10\ M_{\odot}$.

\begin{figure}
    \center
    \includegraphics[width=0.49\textwidth]{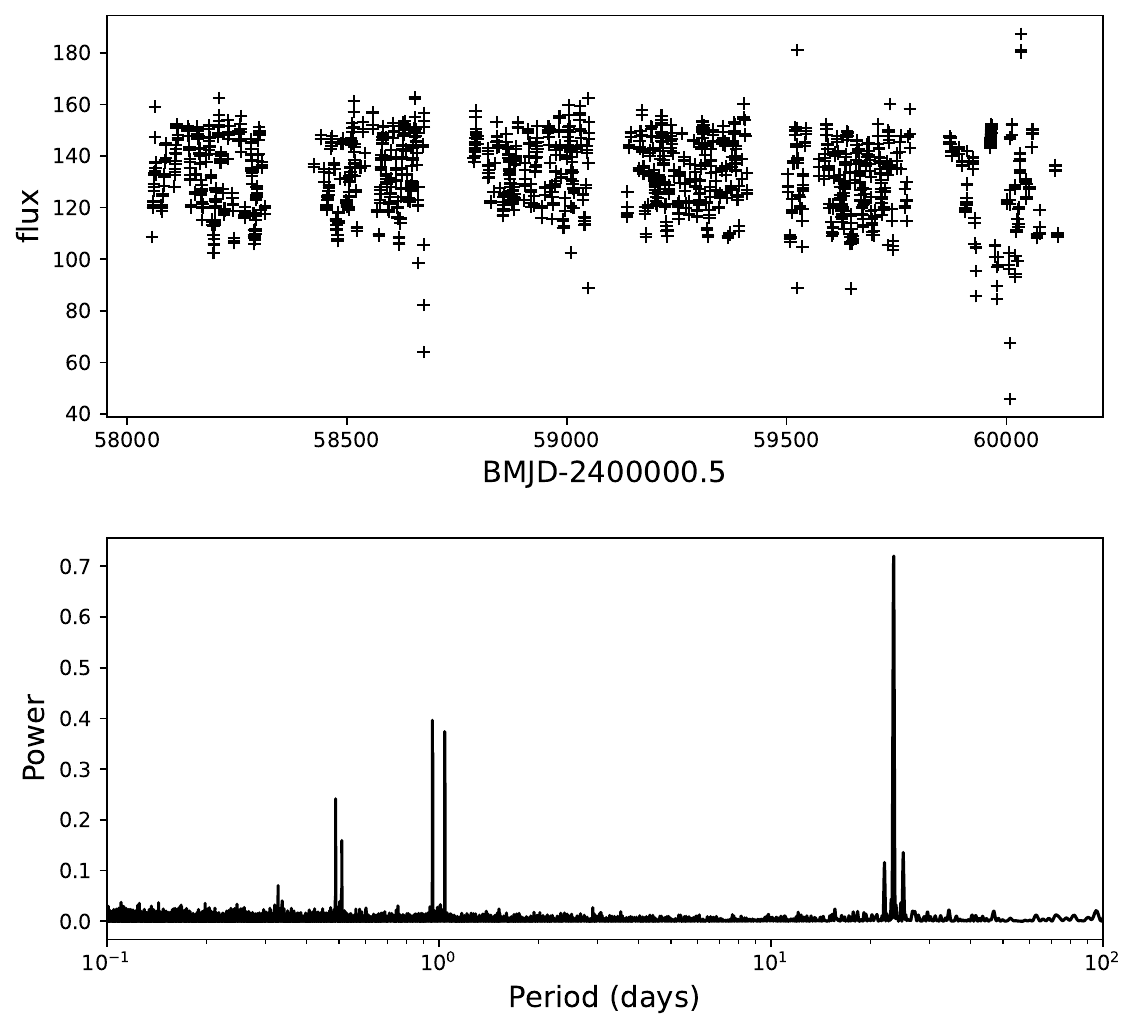}
    \caption{Top panel: ASAS-SN $g$-band light curve. Bottom panel: Lomb–Scargle periodogram. The highest peak marks half of the orbital period from {\it Gaia} DR3.}
    \label{8441_period.fig}
\end{figure}

\begin{figure}
    \center
    \includegraphics[width=0.49\textwidth]{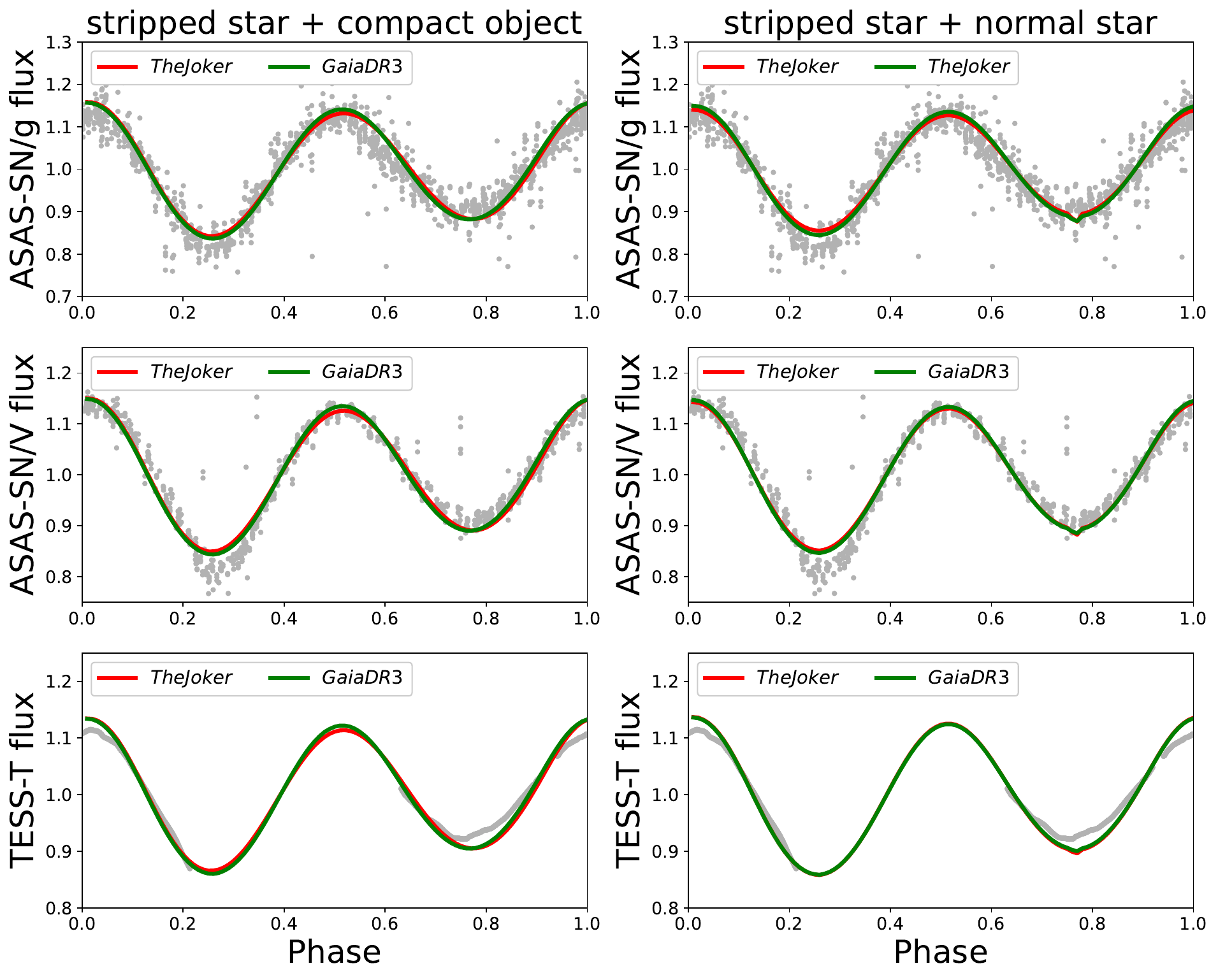}
    \caption{Folded light curves (from top to bottom: ASAS-SN $g$, ASAS-SN $V$, TESS) of G8441 by using the orbital period of 46.895 day. The left panels show the fitting for the scenario of a stripped star and a compact object, while the right panels show the fitting for the scenario of a stripped star and a main-sequence star. 
    The red and green lines represent the best-fitting results from PHOEBE, using the orbital solution from {\it The Joker} and {\it Gaia} DR3, respectively.}
    \label{8441_lcs.fig}
\end{figure}

The Lomb-Scargle algorithm \citep{1976Ap&SS..39..447L,1982ApJ...263..835S} was used to calculate the orbital period ($\approx$46.885 days) with the ASAS-SN $g$-band light curve (Figure \ref{8441_period.fig}).
To obtain a more precise period estimation, we further folded the light curves with periods ranging from 46.5 to 47 days, in a step of 0.001 day.
Through visual inspection of the folded light curves, we determined a period of $\approx$46.895 days.
Figure \ref{8441_lcs.fig} displays the folded light curves in multiple bands.

We employed the software Physics of eclipsing binaries \citep{2016ApJS..227...29P, 2018ApJS..237...26H, 2020ApJS..250...34C} (PHOEBE) to fit the ASAS-SN $V$- and $g$-band light curves.
The ellipsoidal modulation (Figure \ref{8441_lcs.fig}) and the substantial variation amplitude indicate the visible star is a stripped star.
During the LC fitting, we used the {\it eclipse\_method $=$ only\_horizon} as the eclipse model and specified a semi-detached system for the binary configuration.
Two scenarios were considered: one with a stripped star and a compact object, and the other with a stripped star and a standard main-sequence star.
In the first case, a cold ($T_{\rm eff} =$ 300 $K$) and small ($R = 3 \times 10^{-6} R_{\odot}$) blackbody was used as the secondary.
We used two different sets of $a$sin$i$ values: one from {\it The Joker} fitting and the other from the {\it Gaia} DR3 solution (Table \ref{keplerorbit.tab}).
The bolometric gravity darkening coefficients was adopted as $\beta = 0.58$, which was derived from \cite{2011A&A...529A..75C} in the $V$ band for stars with similar atmospheric parameters.
Table \ref{phoebe.tab} presents the parameter estimates from the PHOEBE fitting.
For the scenario of a stripped star and a compact object, the fitting results show that the unseen star is a neutron star or a massive white dwarf, while
for the scenario of a stripped star and a main-sequence star, the best-fit models indicate the unseen dwarf is an A- or F-type star.

\begin{table*}
\caption{Parameter estimates from PHOEBE for G8441. \label{phoebe.tab}}
\centering
\setlength{\tabcolsep}{1.5mm}
 \begin{tabular}{cccccc}
\hline
Binary type & Input & Parameter & System & Primary & Secondary \\
\hline
\multirow{16}*{stripped star + compact object} & \multirow{8}*{{\it The Joker}} & $P_{\rm orb}$ (${\rm d}$) & 46.81 (fixed) \\
& & $e$ & 0.02 (fixed) \\
& & $i$ ($ \ ^{\circ}$) & $68.46^{+0.82}_{-0.19}$ \\
& & $q$ & $0.19^{+0.01}_{-0.01}$ \\
& & $a{\rm sin}i$ (${\rm R}_{\odot}$) & & 51.55 (fixed) & $9.91^{+0.69}_{-0.67}$ \\
& & $R$ (${\rm R}_{\odot}$) & & $16.25^{+0.85}_{-0.84}$ & $3\times10^{-6}$ (fixed) \\
& & $T_{\rm eff}$ (${\rm K}$) & & $4195.0^{+1.35}_{-1.74}$ & 300 (fixed) \\
& & $M$ (${\rm M}_{\odot}$) & & $0.28^{+0.05}_{-0.04}$ & $1.48^{+0.22}_{-0.20}$ \\
\cline{2-6}
& \multirow{8}*{{\it Gaia} DR3} & $P_{\rm orb}$ (${\rm d}$) & 46.88 (fixed) \\
& & $e$ & 0.01 (fixed) \\
& & $i$ ($ \ ^{\circ}$) & $70.15^{+0.63}_{-0.26}$ \\
& & $q$ & $0.23^{+0.01}_{-0.01}$ \\
& & $a{\rm sin}i$ (${\rm R}_{\odot}$) & & 47.91 (fixed) & $11.02^{+0.48}_{-0.47}$ \\
& & $R$ (${\rm R}_{\odot}$) & & $16.19^{+0.34}_{-0.33}$ & $3\times10^{-6}$ (fixed) \\
& & $T_{\rm eff}$ (${\rm K}$) & & $4193.1^{+1.63}_{-0.91}$ & 300 (fixed) \\
& & $M$ (${\rm M}_{\odot}$) & & $0.28^{+0.02}_{-0.02}$ & $1.22^{+0.03}_{-0.03}$ \\
\hline
\multirow{16}*{stripped star + normal star} & \multirow{8}*{{\it The Joker}} & $P_{\rm orb}$ (${\rm d}$) & 46.81 (fixed) \\
& & $e$ & 0.02 (fixed) \\
& & $i$ ($ \ ^{\circ}$) & $75.85^{+0.01}_{-0.04}$ \\ 
& & $q$ & $0.20^{+0.01}_{-0.01}$ \\
& & $a{\rm sin}i$ (${\rm R}_{\odot}$) & & 51.55 (fixed) & $10.31^{+0.71}_{-0.68}$ \\
& & $R$ (${\rm R}_{\odot}$) & & $15.86^{+0.83}_{-0.83}$ & $1.43^{+0.04}_{-0.04}$ \\
& & $T_{\rm eff}$ (${\rm K}$) & & $4200.0^{+0.41}_{-1.13}$ & $8190.2^{+805.7}_{-824.1}$ \\
& & $M$ (${\rm M}_{\odot}$) & & $0.26^{+0.04}_{-0.04}$ & $1.32^{+0.19}_{-0.17}$ \\
\cline{2-6}
& \multirow{8}*{{\it Gaia} DR3} & $P_{\rm orb}$ (${\rm d}$) & 46.88 (fixed) \\
& & $e$ & 0.01 (fixed) \\
& & $i$ ($ \ ^{\circ}$) & $75.17^{+0.02}_{-0.03}$ \\
& & $q$ & $0.24^{+0.01}_{-0.01}$ \\
& & $a{\rm sin}i$ (${\rm R}_{\odot}$) & & 47.91 (fixed) & $11.50^{+0.48}_{-0.48}$ \\
& & $R$ (${\rm R}_{\odot}$) & & $16.07^{+0.31}_{-0.32}$ & $1.13^{+0.01}_{-0.01}$ \\
& & $T_{\rm eff}$ (${\rm K}$) & & $4192.7^{+1.26}_{-0.74}$ & $7195.0^{+135.4}_{-84.8}$ \\
& & $M$ (${\rm M}_{\odot}$) & & $0.28^{+0.02}_{-0.02}$ & $1.15^{+0.02}_{-0.02}$ \\
\hline
\end{tabular}
\end{table*}

\subsection{The nature of the unseen object}
\label{natG8441.sec}

G8441 has been reported as a possible compact object candidate in previous studies \citep{2019ApJ...872L..20G, 2019AJ....158..179Z}.
In this section, we will try to investigate the nature of the unseen object by examining the optical spectra, and the UV and X-ray emission.
Due to the limited number of spectral observations, we did not perform the spectral disentangling.

\begin{figure}
    \center
    \includegraphics[width=0.49\textwidth]{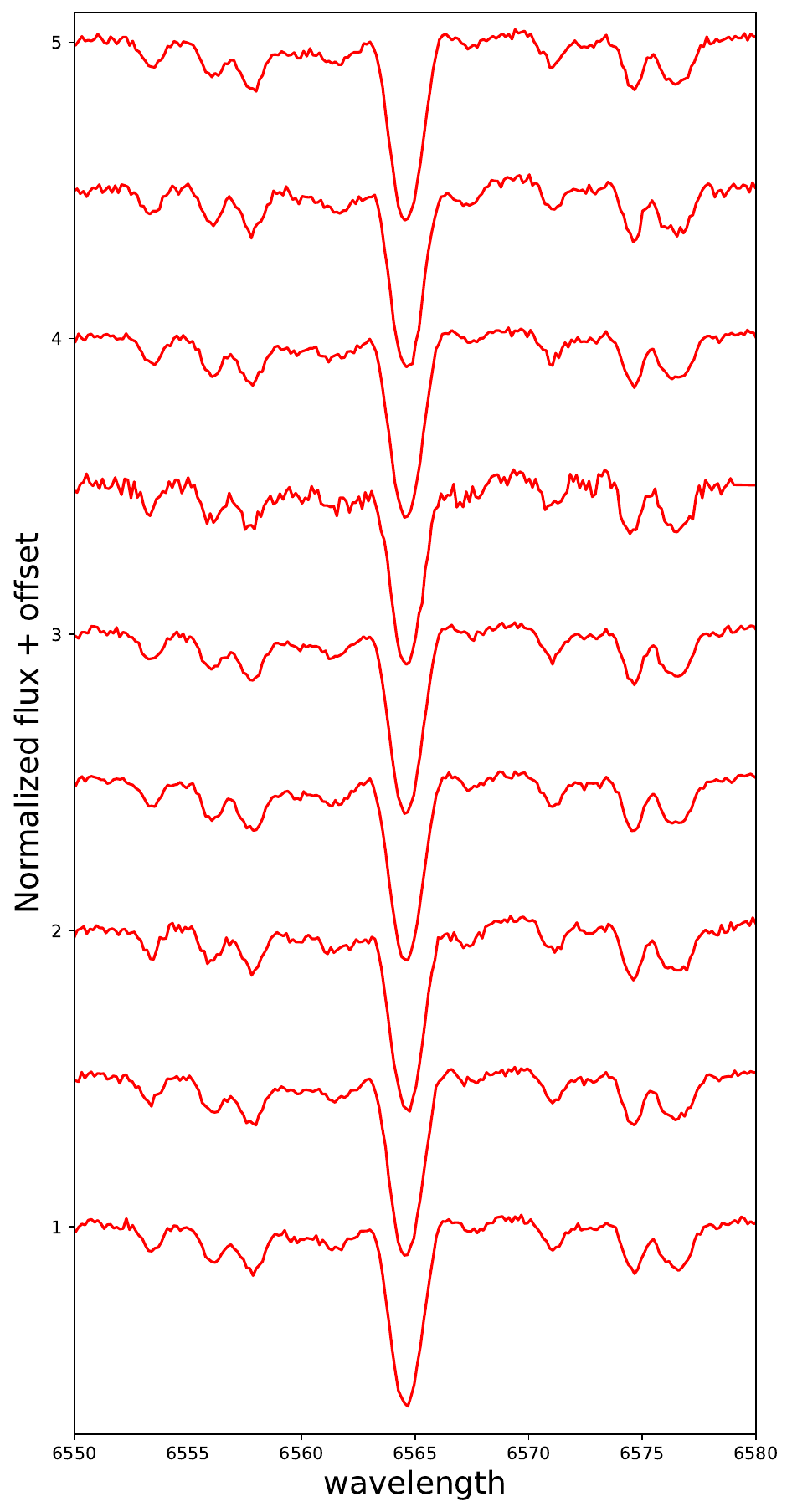}
    \caption{Profiles of the RV-corrected $H_{\alpha}$ lines of G8441.}
    \label{8441_Ha.fig}
\end{figure}

No emission line is detectable in the optical spectra (Figure \ref{HR_diagram.fig}).
It is evident that the $H_{\alpha}$ lines in all spectral observations are absorption lines (Figure \ref{8441_Ha.fig}), indicating the absence of an accretion disk.
This contrasts with most binaries containing a stripped giant star, such as V723 Mon and 2M04123153+6738486 \citep{2021MNRAS.504.2577J,2022MNRAS.516.5945J}, which display clear $H_{\alpha}$ emission lines.
G8441 is likely similar to a few binaries without emission lines, explained by temporary halted or slowed accretion \citep{2022MNRAS.515.1266E}.

\begin{figure}
    \center
    \includegraphics[width=0.47\textwidth]{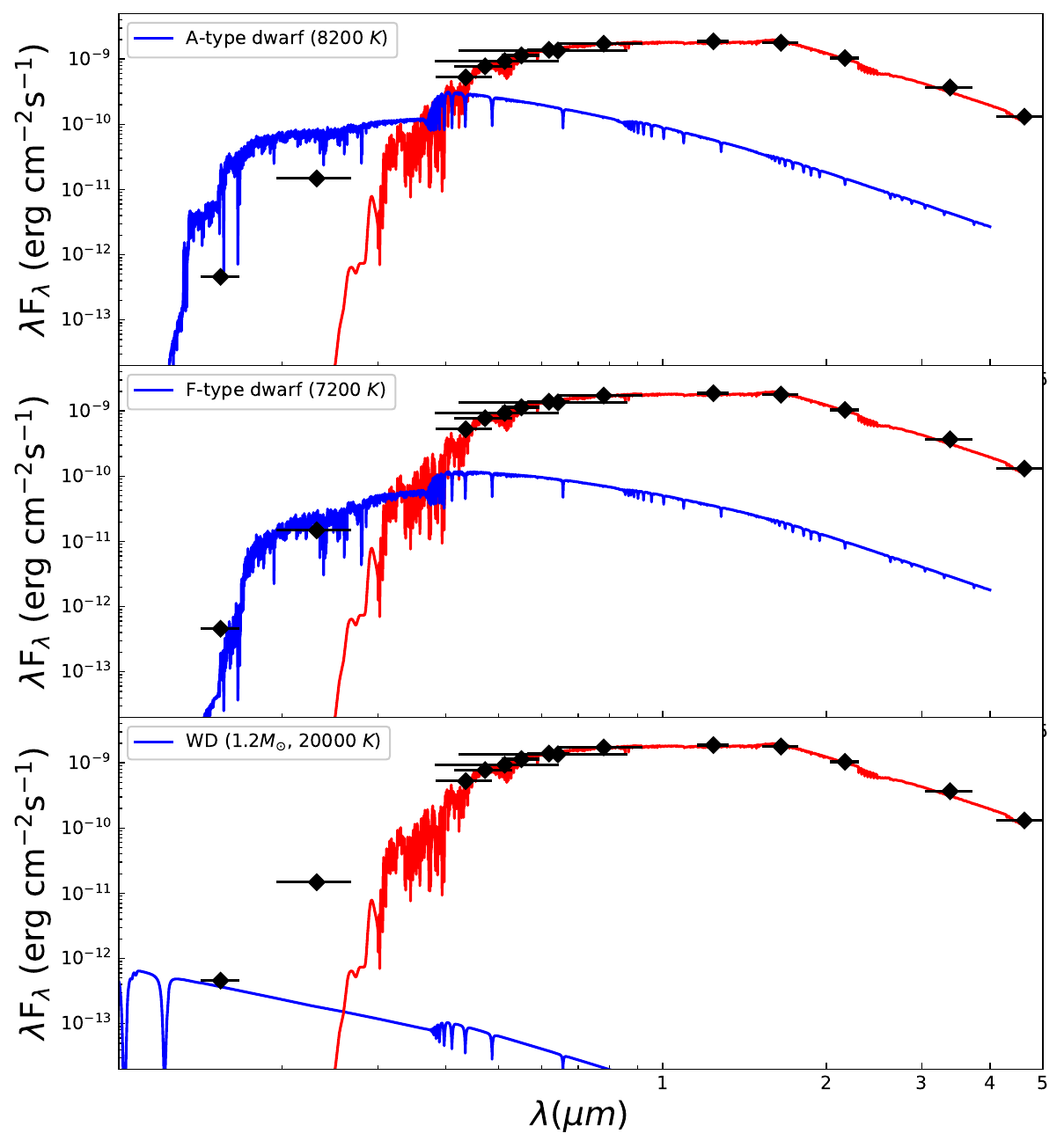}
    \caption{SED of G8441. Top panel: The black points are the multi-band observations of G8441. The red line represents the spectrum of the stripped star, while the blue line marks the spectrum of an A-type main-sequence ($T_{\rm eff} \approx$ 8200 $K$ from PHOEBE fitting). Middle panel: The blue line represents the spectrum of an F-type dwarf ($T_{\rm eff} \approx$ 7200 $K$ from PHOEBE fitting). Bottom panel: The blue line represents the spectrum of a white dwarf with a mass of $1.2 M_{\odot}$ and a temperature of 20,000 $K$.}
    \label{wd_giant.fig}
\end{figure}

G8441 has been observed in FUV and NUV bands by {\it GALEX}.
By using the UV magnitudes of $f_{\rm FUV} = 23.28$ $\mu$Jy and $f_{\rm NUV} = 1142.51$ $\mu$Jy, we calculated the luminosities as $L_{\rm FUV}=1.16 \times 10^{31}$ erg s$^{-1}$ and $L_{\rm NUV}=6.19 \times 10^{32}$ erg s$^{-1}$. 
This luminosity can be attributed to the giant star, given that K-type giants typically exhibit FUV and NUV luminosities within the ranges of $\approx 5\times10^{29}$--$5\times10^{32}$ erg s$^{-1}$ and $\approx 3\times10^{29}$--$7\times10^{33}$ erg s$^{-1}$, respectively \citep{2020ApJ...902..114W}.
However, it is uncertain whether a stripped star has the same magnetic dynamo mechanism as a normal giant.
If the the unseen star is an A-type star ($T_{\rm eff} \approx$ 8200 $K$ from PHOEBE fitting), its FUV and NUV emission would exceed the observed levels significantly (Figure \ref{wd_giant.fig}). Therefore, the A-type star scenario can be ruled out.
If the unseen star is an F-type star ($T_{\rm eff} \approx$ 7200 $K$ from PHOEBE fitting), its UV emission are compatible with the observation.
On the other hand, if the unseen object is a white dwarf, the upper limit of its surface temperature can be constrained with the FUV emission.
White dwarfs with effective temperatures greater than 20,000 $K$ can be ruled out since models predict higher fluxes than the {\it GALEX} FUV observation (Figure \ref{wd_giant.fig}). 
Thus, the UV luminosities (especially the high NUV luminosity) can be either totally from the stripped star, or from a combination of the stripped star and the unseen object.

G8441 was observed by {\it XMM-Newton} on May 4, 2020.
No X-ray counterpart was detected from the EPIC instruments, including M1, M2, and PN.
Using a radius of 10$^{\prime\prime}$, an upper limit of the net count rate in the 0.3–-8 keV range was calculated with the PN instrument, yielding approximately 0.0004 counts/sec.
We utilized the PIMMS tool\footnote{https://heasarc.gsfc.nasa.gov/cgi-bin/Tools/w3pimms/w3pimms.pl} to convert the count rate into flux.
Applying the standard linear relation between the hydrogen column density $N_{\rm H}$ and the reddening $N_{\rm H} = 5.8\times10^{21}/E(B-V)$ \citep{1998ApJ...500..525S}, the hydrogen column density $N_{\rm H}$ was calculated to be $5.8\times10^{19}$ cm$^{-2}$ with the optical reddening of $E(B-V) = 0.01$ mag.
By using a power-law photon index of 1.5 or 2, the upper limit of the unabsorbed flux in the 0.3--8 keV band was derived to be $1.1\times10^{-15}$ erg cm$^{-2}$ s$^{-1}$ or $7.9\times10^{-16}$ erg cm$^{-2}$ s$^{-1}$, respectively.
At the adopted distance of 857 pc, the upper limit of the emitted luminosity was estimated to be $L_{X} \lesssim$ $9.7\times10^{28}$ erg s$^{-1}$ or $6.9\times10^{28}$ erg s$^{-1}$.

To sum up, the substantial variation amplitude of light curves indicates the visible star is most likely a stripped star, although there is no emission feature in the optical spectra.
The unseen object could be a compact object (white dwarf or neutron star) or a main-sequence F star.
Unfortunately, present observational data (e.g., UV and X-ray data, SED, and optical spectra) do not provide sufficient evidence to distinguish between these scenarios.

\subsection{Evolutionary track}

The Binary Population And Spectral Synthesis code (BPASS) \citep{bpass2017,10.1093/mnras/sty1353} is a tool to calculate the evolutionary tracks of single stars or binary systems.
We used the hoki package \citep{2020JOSS....5.1987S} to search for binary models consistent with the observed properties of G8441 based on the results from BPASS.
The selection criteria included $P_{\rm orb}=46.88 \pm 5$ days, $V=0.935\pm1$ mag, $J=-1.385\pm1$ mag, $K=-2.23\pm1$ mag, $\log T_{\rm eff} (K)=3.62\pm0.04$, $M_{1}=0.28\pm0.1\ M_{\odot}$, and $M_{2}$ in the range of 1.1--1.5 $M_{\odot}$, where $M_{1}$ and $M_{2}$ represent the masses of the giant and the invisible star, respectively.
We identified four systems including an F- or G-type star and one system with a compact object (Table \ref{bpass.tab}).
The parameters of these systems are not entirely consistent with the analysis in Section \ref{G8441orbit} and \ref{natG8441.sec}.
For example, the masses of the stripped star in these models are higher than our calculation ($\approx$0.28 $M_{\odot}$), and for the four models containing normal stars, the temperatures of these normal star are lower than the PHOEBE fitting results (Table \ref{phoebe.tab}).

One interesting finding is that the mass of the stripped star ($\approx$0.28 $M_{\odot}$) follows the relationship between the mass and orbital period 
($M_{\rm WD}/M_{\odot} = (\frac{P_{\rm orb}/{\rm day}}{1.1\times10^{5}})^{1/4.75} + 0.115$) 
of white dwarfs or the core mass of low-mass giants \citep{1999A&A...350..928T}. 
This suggests that the mass transfer in G8441 might have been close to termination and most of the hydrogen envelope of the stripped star has likely been stripped away.
This donor star could eventually evolve into an extremely low mass white dwarf.

\begin{table*}
\caption{Possible binary properties of G8441 by matching the BPASS v2.2.1 models at ${\rm Z =0.002}$.\label{bpass.tab}}
\centering
\setlength{\tabcolsep}{10pt}
\begin{tabular}{cccccccc}
\hline\noalign{\smallskip}
$M_{1}$ ($M_{\odot}$) & $T_{\rm eff,1}$ ($K$) & $R_{1}$ ($R_{\odot}$) & $M_{2}$ ($M_{\odot}$) & $T_{\rm eff,2}$ ($K$) & $R_{2}$ ($R_{\odot}$) & $q$ & Period (day) \\
\hline\noalign{\smallskip}
0.37 & 4543.1 & 17.8 & 1.18 & 5434.0 & 1.08 & 0.314 & 42.758 \\
0.35 & 4533.0 & 17.6 & 1.21 & 5251.1 & 0.95 & 0.291 & 43.809 \\
0.35 & 4517.9 & 18.1 & 1.33 & 5539.6 & 0.98 & 0.266 & 45.566 \\
0.35 & 4516.8 & 18.3 & 1.41 & 5712.8 & 0.99 & 0.247 & 46.393 \\
0.36 & 4511.0 & 18.5 & 1.47 & - & - & 0.244 & 46.851 \\   
\hline\noalign{\smallskip}
\end{tabular}
\end{table*}

\section{Spatial distribution in the Milky Way}

Until now, about ten binaries including white dwarfs or neutron stars have been discovered by RV.
We collected these binary systems \citep{2022MNRAS.517.4005M,2022ApJ...940..165Y,2022ApJ...938...78L,2022NatAs...6.1203Y,2022ApJ...933..193Z,2022arXiv221004685Z,2022ApJ...936...33Z} to investigate their positions in the Milky Way.

\begin{figure*}
    \center
    \includegraphics[width=1\textwidth]{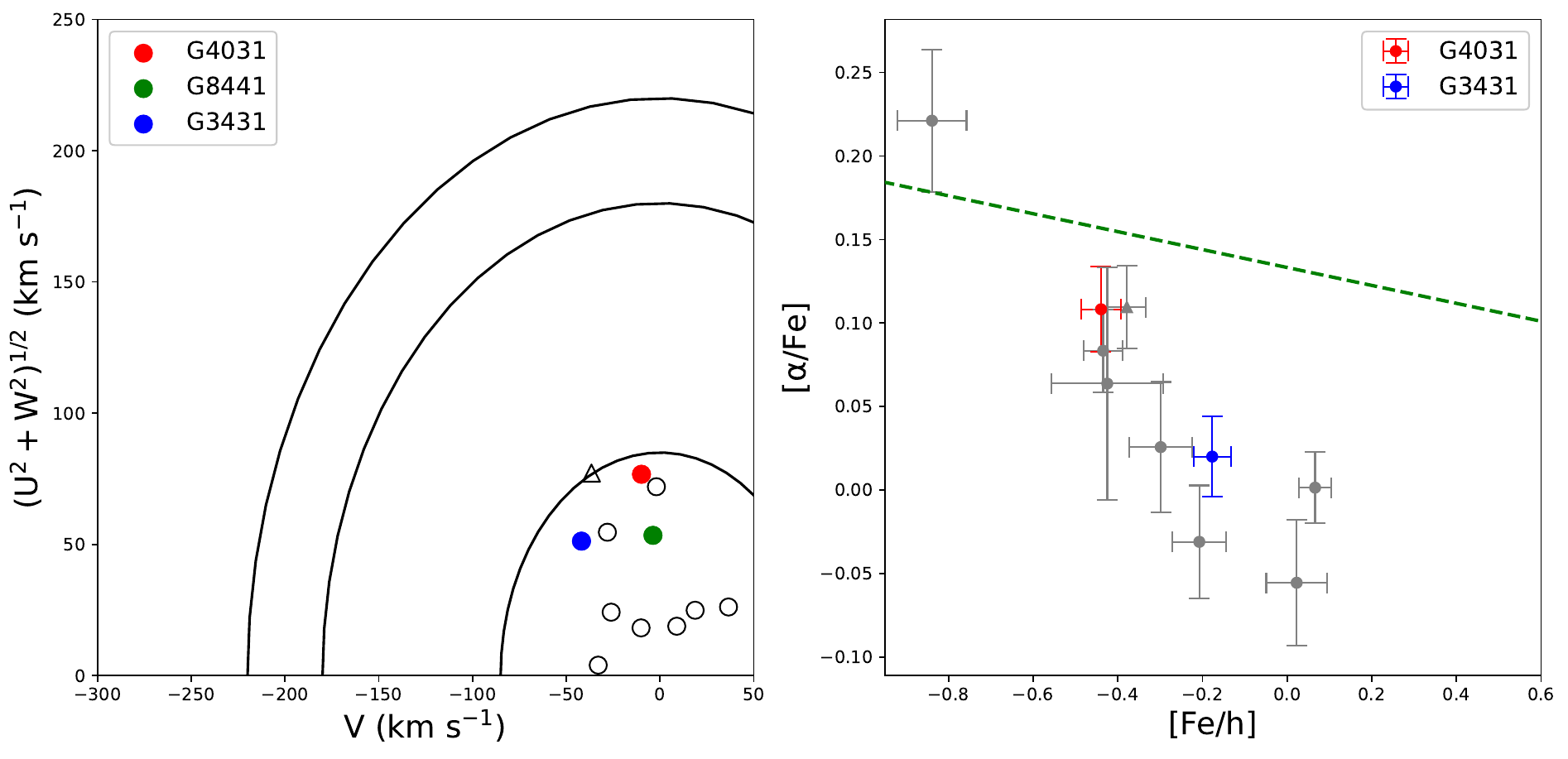}
    \caption{Spatial distribution of binaries including white dwarfs or neutron stars. Left panel: The Toorme diagram of the binary systems. The circles and triangles represents stars classified as in the thin disk or thick disk using the kinematic method. Right panel: The ${\rm [Fe/H]}$ - ${\rm [\alpha/Fe]}$ diagram of the binary systems. The green dashed line indicates the approximate border of thin disk and thick disk from metallicity \citep{2015MNRAS.453.1855M}.}
    \label{toomre.fig}
\end{figure*}

To identify whether one source belongs to the thin disk, thick disk, or halo,
first, we defined one probability assuming that the velocities
($U_{LSR}$, $V_{LSR}$, $W_{LSR}$) follow a 3-D Gaussian distribution \citep{2016RAA....16...44G}:
\begin{equation}
{\rm Prob} = c{\cdot}{\rm exp}\{-\frac{U_{\rm LSR}^2}{2\sigma_{U}^2} -\frac{(V_{\rm LSR}- V_{\rm asym})^2}{2\sigma_{V}^2} -\frac{W_{\rm LSR}^2}{2\sigma_{W}^2} \},
\end{equation}
where $c = (2\pi)^{-3/2}(\sigma_{U}\sigma_{V}\sigma_{W})^{-1}$ normalizes the expression.
Here $V_{\rm asym}$ is the asymmetric drift, and $\sigma_{U}$, $\sigma_{V}$, and $\sigma_{W}$ are the velocity dispersions in three dimensions, all of which vary for different components\footnote{$(U,V,W)_{\odot, \rm LSR} = (11.1, 12.24, 7.25) $ km/s; \\
Thin disk: $(V_{\rm asym}, \sigma_U, \sigma_V, \sigma_W) = (-12, 39, 20, 20) $ km/s; \\
Thick disk: $(V_{\rm asym}, \sigma_U, \sigma_V, \sigma_W) = (-51, 63, 39, 39) $ km/s; \\
Halo: $(V_{\rm asym}, \sigma_U, \sigma_V, \sigma_W) = (-199, 141, 106, 94) $ km/s.}.
Second, we defined the probability ratio of belonging to different components as
\begin{equation}
f_{\rm thin/thick/halo} = \frac{\rm Prob_{thin/thick/halo}}{\rm Prob_{thin} + Prob_{thick}+Prob_{halo}}.
\end{equation}
One star was classified as a candidate in the thin disk, thick disk, or halo when the corresponding ratio exceeded 50\%.
Figure \ref{toomre.fig} (left panel) shows the Toorme diagram of those binary systems.
Furthermore, we collected the [$\alpha$/Fe] and [Fe/H] values from \cite{2022ApJS..259...51W} and segregated these systems into the thin disk and thick disk based on the [$\alpha$/Fe]--[Fe/H] diagram (Figure \ref{toomre.fig}).
Both the kinematic and chemical methods suggest most of these systems belong to the Galactic thin disk (Table \ref{uvw.tab}).

\begin{table*}
\caption{Spatial distribution parameters of binaries including white dwarfs or neutron stars.\label{uvw.tab}}
\begin{center}
\setlength{\tabcolsep}{7pt}
\begin{tabular}{ccccccccccc}
\hline\noalign{\smallskip}
{\it Gaia} ID & R.A. ($^{\circ}$) & Decl. ($^{\circ}$) & $v_{0}$ (km/s) & D (pc) & ($U,V,W$)${\rm _{LSR}}$ & ${\rm [Fe/H]}$ & ${\rm [\alpha/Fe]}$ & Classification$^a$ \\
\hline\noalign{\smallskip}
4031997035561149824 & 176.46978 & 35.29075 & 35.7 & 719.8 & (-74.1, -9.9, 19.8) & -0.440 & 0.108 & thin/thin \\
3431326755205579264 & 90.61903 & 28.13287 & 63.7 & 338.8 & (-51.0,-41.9,-5.3) & -0.177 & 0.020 & thin/thin \\
844176650958726144 & 169.12464 & 55.72840 & 20.5 & 858.8 & (-53.4,-3.8,1.8) & - & - & thin/- \\
1577114915964797184 & 193.98563 & 56.97958 & -1.7 & 577.8 & (-23.5,18.7,-8.3) & -0.435 & 0.083 & thin/thin \\
3425175331243738240 & 94.14802 & 23.31924 & 29.0 & 1037.7 & (-15.9,-10.1,-8.8) & 0.066 & 0.001 & thin/thin \\
2874966759081257728 & 358.73652 & 33.94047 & 41.0 & 127.3 & (-9.1,36.5,-24.5) & -0.299 & 0.026 & thin/thin \\
608426858154004864 & 131.96670 & 13.45494 & 87.5 & 1092.5 & (-76.1,-36.6,12.6) & -0.379 & 0.110 & thick/thin \\
608189290627289856 & 133.26905 & 13.34226 & 66.1 & 324.6 & (-53.4,-28.1,11.4) & -0.208 & -0.031 & thin/thin \\
64055043471903616 & 57.50692 & 22.31625 & -17.4 & 2006.8 & (23.5,-26.1,5.9) & 0.023 & -0.056 & thin/thin \\
770431444010267392 & 170.77878 & 40.12676 & -8.0 & 314.5 & (-1.2,-33.0,-3.8) & -0.425 & 0.064 & thin/thin \\
1633051023841345280 & 262.25073 & 65.49800 & -4.2 & 229.4 & (-18.8,8.9,0.3) & - & - & thin/- \\
3298897073626626048 & 64.83362 & 7.42928 & 86.0 & 672.7 & (-65.1,-1.9,-30.7) & -0.839 & 0.221 & thin/thick \\
\hline\noalign{\smallskip}
\end{tabular}
\end{center}
\footnotesize{$^a$ The classification is from the kinematic and chemical methods.}
\end{table*}

\section{Summary}
\label{sum.sec}

By combining the {\it Gaia} DR3 astrometric solution and LAMOST DR9 LRS and MRS data, we discover three binaries with possible compact components (i.e., G4031, G3431 and G8441). 

G4031 is a binary system with a period of $P \approx$ 140 day consisting of a G-type main-sequence star. 
The evolutionary and gravitational masses of the visible star are estimated to be $1.46^{+0.02}_{-0.02} M_{\odot}$ and $1.62\pm0.02 M_{\odot}$, respectively.
The Kepler solution from {\it The Joker} matches better with the RV data than the orbit from {\it Gaia} DR3.
By using {\it The Joker} results, the mass function is calculated to be $f(M) \approx 0.092\pm0.006 M_{\odot}$. 
These lead to a minimum mass of the unseen star of $0.77\pm0.03 M_{\odot}$ (calculated with the evolutionary mass of the visible star) or $0.82\pm0.02 M_{\odot}$ (calculated with the gravitational mass of the visible star), suggesting the unseen star is likely a white dwarf.

G3431 has a orbital period of $P \approx$ 120 day containing a G-type main-sequence star. 
Using stellar evolution models, the mass of the visible star is determined to be $M=1.42^{+0.03}_{-0.03} M_{\odot}$.
The orbit solutions obtained from {\it The Joker} and {\it Gaia} DR3 are in good agreement, indicating a mass function of $f(M) \approx 0.324\pm0.038 M_{\odot}$ or $0.392\pm0.023 M_{\odot}$, respectively. 
Therefore, an invisible star with a minimum mass of $1.36\pm0.09 M_{\odot}$ or $1.49\pm0.05 M_{\odot}$ is obtained, suggesting the presence of a white dwarf or neutron star.

G8441 is a binary with a period of $P \approx$ 46.895 day. 
Different with G4031 and G3431, the light curves of G8441 strongly suggest that the visible star is a stripped star.
Using the period derived from {\it Gaia} DR3, we calculated the mass of the stripped star to be $0.28\pm0.03\ M_{\odot}$.
The RV fitting indicates a mass function of $f(M) \approx 0.84\pm0.12 M_{\odot}$ and a minimum mass of $1.25\pm0.17 M_{\odot}$ for the unseen star.
By fitting the multi-band light curves, we determined the orbital inclination angle to be around 70 degrees.
The mass range of the invisible star is 1.1--1.5 $M_{\odot}$, implying it is possibly a compact object (white dwarf or neutron star) or an F-type main-sequence star.
Although the visible star is recognized as a stripped star, there are no detectable emission lines in the optical spectra.
This intriguing feature may be attributed to a temporary interruption or reduction in the accretion process.

For the two sources with a main-sequence companion, by examining the {\it Gaia} XP spectra and applying spectral disentangling to the LAMOST MRS observations, we have not identified any additional optical component in addition to the visible star, supporting the scenario that these sources are binary systems with compact components.
An examination of about ten binaries containing white dwarfs or neutron stars using both the kinematic and chemical methods suggest most of these systems are located in the Galactic thin disk.

RV and astrometry have opened up new possibilities for searching for binaries with compact components. 
In addition to LAMOST, the RV data from other large spectroscopic surveys, including APOGEE, Radial Velocity Experiment (RAVE), GALactic Archaeology with HERMES (GALAH), and Dark Energy Spectroscopic Instrument (DESI), should also be collected and utilized.
By combining the epoch RV data and astrometric data from {\it Gaia}, particularly with epoch astrometric measurements from {\it Gaia} DR4 in the future, numerous compact objects hidden in wide binaries can be discovered. 
A comprehensive sample of compact objects with accurate dynamical mass measurements will significantly contributes to building the mass function of compact objects.
Especially, a sample of massive neutron stars and low-mass black holes can help confirm the maximum mass of neutron stars and minimum mass of black holes, and help solve the mass gap problem---the scarcity of black holes with masses ranging from 2.5 to 5 $M_{\odot}$ \citep{1998ApJ...499..367B}.
These discoveries, along with the compact objects found by other methods (e.g., X-ray, gravitational wave, and gravitational microlensing), are expected to provide valuable insights into the late evolution of massive stars (in binary systems) and help us better understand the supernova mechanism and the black hole formation process.

\begin{acknowledgements}

We thank the anonymous referee for helpful comments and suggestions that have improved the paper. 
We thank Dr. Kareem El-Badry for very useful discussion about binaries including stripped stars.
The Guoshoujing Telescope (the Large Sky Area Multi-Object Fiber Spectroscopic Telescope LAMOST) is a National Major Scientific Project built by the Chinese Academy of Sciences. Funding for the project has been provided by the National Development and Reform Commission. LAMOST is operated and managed by the National Astronomical Observatories, Chinese Academy of Sciences. 
This work uses data obtained through the Telescope Access Program (TAP), which has been funded by the TAP member institutes. 
This work presents results from the European Space Agency (ESA) space mission {\it Gaia}. {\it Gaia} data are being processed by the {\it Gaia} Data Processing and Analysis Consortium (DPAC). Funding for the DPAC is provided by national institutions, in particular the institutions participating in the {\it Gaia} MultiLateral Agreement (MLA). The {\it Gaia} mission website is https://www.cosmos.esa.int/gaia. The {\it Gaia} archive website is https://archives.esac.esa.int/gaia. We acknowledge use of the VizieR catalog access tool, operated at CDS, Strasbourg, France, and of Astropy, a community-developed core Python package for Astronomy (Astropy Collaboration, 2013). This research made use of Photutils (Bradley et al. 2020), an Astropy package for detection and photometry of astronomical sources. This work was supported by National Science Foundation of China (NSFC) under grant Nos. 11988101/11933004/11833002/12090042/12103047, National Key Research and Development Program of China (NKRDPC) under grant Nos. 2019YFA0405504 and 2019YFA0405000, and Strategic Priority Program of the Chinese Academy of Sciences under
grant No. XDB41000000. S.W. acknowledges support from the Youth Innovation Promotion Association of the CAS (IDs 2019057).

\end{acknowledgements}

\bibliographystyle{aasjournal}
\bibliography{bibtex.bib}{}

\clearpage
\appendix
\renewcommand*\thetable{\Alph{section}.\arabic{table}}
\renewcommand*\thefigure{\Alph{section}\arabic{figure}}

\section{RV measurements of our targets}
\label{rvdata_appendix.sec}

\setcounter{table}{0}

 \begin{longtable}{cccccc|ccccc}
 \caption{Barycentric-corrected RV values of our targets. \label{lamost3.tab}}
 \\\hline\noalign{\smallskip}
 name & BMJD & RV & Uncertainty & {\it SNR} & Resolution & BMJD & RV & Uncertainty & {\it SNR} & Resolution\\
 & (day) & (km/s) & (km/s) &  &  & (day) & (km/s) & (km/s) & & \\
 \hline\noalign{\smallskip}
 \multirow{35}*{G4031} & 58495.85485 & 19.76 & 1.08 & 80.67 & MRS & 58863.84738 & 35.27 & 1.49 & 25.44 & MRS \\
& 58495.87082 & 17.76 & 2.24 & 45.8 & MRS & 58863.86336 & 34.77 & 1.35 & 24.8 & MRS \\
& 58495.88679 & 19.26 & 2.24 & 42.04 & MRS & 58883.76181 & 22.76 & 1.12 & 54.22 & MRS \\
& 58495.90346 & 17.26 & 2.24 & 38.31 & MRS & 58883.77778 & 22.26 & 2.24 & 51.46 & MRS \\
& 58495.91943 & 18.26 & 1.10 & 52.18 & MRS & 58883.79375 & 22.76 & 2.24 & 50.61 & MRS \\
& 58511.81416 & 29.27 & 2.24 & 70.09 & MRS & 58886.77310 & 21.26 & 2.24 & 20.91 & MRS \\
& 58511.83014 & 25.27 & 2.24 & 68.23 & MRS & 58886.78907 & 25.27 & 2.24 & 20.63 & MRS \\
& 58511.84611 & 31.27 & 2.24 & 62.42 & MRS & 58886.80574 & 21.26 & 2.24 & 20.42 & MRS \\
& 58535.73989 & 42.78 & 1.10 & 47.15 & MRS & 58895.74655 & 19.76 & 2.24 & 48.55 & MRS \\
& 58541.74577 & 43.78 & 2.24 & 46.75 & MRS & 58895.76252 & 19.26 & 2.24 & 43.29 & MRS \\
& 58541.76244 & 44.28 & 1.10 & 64.59 & MRS & 58895.77849 & 19.76 & 2.24 & 40.26 & MRS \\
& 58541.77841 & 47.78 & 1.12 & 64.88 & MRS & 58911.68074 & 18.76 & 1.08 & 65.47 & MRS \\
& 58541.79438 & 46.78 & 1.10 & 63.2 & MRS & 58911.69671 & 19.76 & 1.10 & 57.71 & MRS \\
& 58557.69929 & 56.29 & 1.08 & 69.1 & MRS & 58912.69884 & 19.26 & 2.24 & 15.36 & MRS \\
& 58557.71665 & 56.79 & 1.10 & 66.93 & MRS & 58912.71482 & 19.76 & 2.24 & 11.38 & MRS \\
& 58557.73332 & 55.78 & 2.24 & 47.59 & MRS & 58912.73148 & 19.76 & 2.24 & 8.06 & MRS \\
& 58557.74930 & 57.29 & 2.24 & 47.87 & MRS & 58918.70607 & 21.76 & 2.24 & 7.08 & MRS \\
& 58567.63994 & 51.28 & 2.28 & 10.38 & MRS & 58918.72273 & 21.76 & 2.24 & 7.31 & MRS \\
& 58567.65660 & 48.28 & 1.97 & 11.46 & MRS & 58918.74218 & 21.76 & 1.97 & 11.47 & MRS \\
& 58567.67258 & 51.28 & 2.51 & 11.49 & MRS & 58921.68189 & 22.76 & 1.12 & 42.47 & MRS \\
& 58567.68855 & 54.28 & 2.24 & 44.33 & MRS & 58921.69786 & 23.26 & 2.24 & 7.34 & MRS \\
& 58567.70522 & 53.78 & 2.24 & 46.27 & MRS & 58921.71383 & 23.26 & 2.24 & 7.96 & MRS \\
& 58567.72119 & 53.78 & 2.24 & 46.16 & MRS & 58942.63605 & 34.77 & 1.08 & 65.33 & MRS \\
& 58567.73716 & 50.78 & 1.10 & 61.99 & MRS & 58942.65480 & 34.27 & 1.08 & 66.95 & MRS \\
& 58851.88626 & 47.28 & 2.24 & 7.39 & MRS & 58942.67147 & 34.27 & 2.24 & 66.29 & MRS \\
& 58851.90292 & 47.28 & 2.24 & 6.08 & MRS & 58951.61955 & 40.28 & 1.08 & 88.57 & MRS \\
& 58851.91890 & 46.78 & 3.93 & 7.67 & MRS & 58951.63552 & 40.78 & 1.08 & 92.65 & MRS \\
& 58852.87729 & 45.78 & 1.35 & 24.30 & MRS & 58951.65219 & 39.77 & 1.08 & 89.08 & MRS \\
& 58852.89326 & 45.78 & 1.28 & 27.91 & MRS & 58969.51895 & 58.29 & 2.24 & 52.13 & MRS \\
& 58852.90993 & 46.28 & 2.24 & 7.33 & MRS & 58969.53492 & 57.79 & 2.24 & 53.70 & MRS \\
& 58852.92590 & 45.78 & 2.24 & 6.59 & MRS & 58969.55158 & 57.79 & 1.08 & 72.73 & MRS \\
& 58858.84083 & 39.27 & 3.69 & 8.56 & MRS & 58981.54521 & 53.78 & 5.49 & 64.98 & MRS \\
& 58858.85680 & 39.27 & 2.55 & 17.91 & MRS & 58981.56188 & 53.78 & 1.10 & 64.59 & MRS \\
& 58858.87347 & 39.27 & 2.42 & 13.16 & MRS & 58981.57785 & 53.78 & 2.24 & 57.89 & MRS \\
& 58863.83072 & 35.27 & 2.24 & 23.36 & MRS & 57442.76677 & 48.28 & 2.47 & 96.51 & LRS \\
\hline
\multirow{21}*{G3431} & 58059.85241 & 90.81 & 2.88 & 181.19 & MRS & 58831.70298 & 68.29 & 2.24 & 48.14 & MRS \\
& 58531.49069 & 75.80 & 2.24 & 6.42 & MRS & 58831.70298 & 74.30 & 1.08 & 126.99 & MRS \\
& 58531.50666 & 73.30 & 2.24 & 5.73 & MRS & 58863.63865 & 80.80 & 2.24 & 158.21 & MRS \\ 
& 58531.50666 & 68.29 & 1.08 & 100.55 & MRS & 58863.63865 & 81.30 & 1.84 & 176.55 & MRS \\
& 58531.52332 & 73.30 & 2.24 & 7.36 & MRS & 58863.65532 & 80.80 & 1.08 & 167.16 & MRS \\
& 58531.53930 & 68.79 & 1.08 & 108.81 & MRS & 58891.57097 & 30.77 & 1.08 & 153.83 & MRS \\
& 58531.55596 & 68.29 & 2.24 & 130.21 & MRS & 58891.57097 & 31.27 & 2.24 & 153.89 & MRS \\
& 58531.57193 & 68.29 & 1.08 & 97.02 & MRS & 59234.54116 & 57.79 & 1.08 & 18.16 & MRS \\
& 58531.58860 & 68.29 & 2.24 & 107.65 & MRS & 59234.54116 & 57.29 & 2.24 & 15.46 & MRS \\
& 58535.47923 & 74.30 & 2.24 & 142.96 & MRS & 59234.57311 & 59.29 & 2.24 & 16.43 & MRS \\
& 58535.47923 & 73.80 & 1.08 & 116.13 & MRS & 59234.57311 & 81.80 & 1.06 & 7.25 & MRS \\
& 58535.51187 & 73.80 & 2.24 & 160.19 & MRS & 59244.53990 & 66.79 & 1.31 & 125.28 & MRS \\
& 58535.52784 & 74.30 & 1.08 & 65.25 & MRS & 59244.55587 & 66.29 & 1.28 & 153.01 & MRS \\
& 58535.52784 & 74.30 & 2.24 & 135.54 & MRS & 59244.55587 & 30.77 & 2.24 & 148.61 & MRS \\
& 58535.56048 & 90.31 & 3.93 & 180.18 & MRS & 59247.59317 & 69.79 & 1.60 & 116.31 & MRS \\
& 58535.56048 & 74.80 & 2.24 & 125.49 & MRS & 59247.60984 & 69.79 & 2.24 & 128.88 & MRS \\
& 58535.57714 & 90.81 & 1.10 & 162.69 & MRS & 59247.60984 & 69.79 & 1.49 & 117.03 & MRS \\
& 58805.75413 & 39.27 & 2.24 & 141.39 & MRS & 59265.47927 & 51.28 & 2.19 & 22.59 & MRS \\
& 58805.75413 & 39.27 & 1.08 & 136.27 & MRS & 55870.00454 & 51.57 & 2.87 & 289.30 & LRS \\
& 58805.78955 & 39.27 & 1.22 & 120.49 & MRS & 56349.52170 & 48.78 & 4.85 & 147.94 & LRS \\
& 58822.72570 & 52.78 & 1.08 & 24.83 & MRS & 56683.61161 & 24.77 & 5.13 & 12.73 & 2.16 m \\
& 58822.74584 & 52.28 & 2.24 & 24.15 & MRS & 59856.82953 & 57.56 & 5.00 & 12.99 & 2.16 m \\
& 58831.68631 & 68.79 & 2.24 & 78.5 & MRS&   & &  & &  \\
\hline
\multirow{6}*{G8441} & 58508.80531 & 64.29 & 1.10 & 39.85 & MRS & 57738.88766 & -34.77 & 4.88 & 152.94 & LRS \\
& 58508.82128 & 63.29 & 1.10 & 27.38 & MRS & 58534.82735 & -47.74 & 1.37 & 8.77 & 2.16 m \\
& 58508.83726 & 64.79 & 1.14 & 22.08 & MRS & 58542.64318 & -6.98 & 8.10 & 40.76 & 2.16 m \\
& 59216.8379 & 71.79 & 1.08 & 135.02 & MRS & 58553.64598 & 44.80 & 12.45 & 48.65 & 2.16 m \\
& 59216.85387 & 71.79 & 1.08 & 122.25 & MRS & 58557.61568 & 60.87 & 3.43 & 50.18 & 2.16 m \\
& 59216.87054 & 71.79 & 1.08 & 137.84 & MRS & 58570.75230 & 18.40 & 2.83 & 41.04 & 2.16 m \\
& 59216.88721 & 71.79 & 1.06 & 135.83 & MRS & 58575.50123 & -44.37 & 1.59 & 47.63 & 2.16 m \\
& 59216.90318 & 71.79 & 1.08 & 127.15 & MRS & 58584.53500 & -38.52 & 5.08 & 43.90 & 2.16 m \\
& 59216.91916 & 71.79 & 1.08 & 119.38 & MRS & 58594.69260 & 19.36 & 4.32 & 32.32 & 2.16 m \\
& 59241.74912 & -28.27 & 1.08 & 95.97 & MRS & 58556.32867 & 76.01 & 0.38 & 38.41 & P200 \\
& 59241.76510 & -27.77 & 1.10 & 38.47 & MRS & 58556.33929 & 74.69 & 0.37 & 38.47 & P200 \\
& 59241.78107 &  -27.77 & 1.06 & 75.13 & MRS & 58556.36981 & 69.91 & 0.38 & 36.96 & P200 \\
& 55975.76376 & 57.29 & 6.76 & 33.46 & LRS & 58556.37701 & 69.19 & 0.38 & 37.51 & P200 \\
& 57013.93101 & 61.79 & 5.20 & 141.46 & LRS &  &  &  &  &  \\
\noalign{\smallskip}\hline
\end{longtable}
\clearpage

\end{document}